\documentclass[aps,preprint,amssymb,showpacs,eqsecnum,tightenlines]{revtex4}

\usepackage{amsmath}
\usepackage{amssymb}
\usepackage{amsthm}
\usepackage{pstricks,pst-node,pst-plot}
\newpsobject{showgrid}{psgrid}{subgriddiv=1, griddots=5, gridlabels=6pt}
\usepackage{graphicx}

\renewcommand{\(}{\left(}
\renewcommand{\)}{\right)}
\renewcommand{\[}{\left[}
\renewcommand{\]}{\right]}

\newcommand{\beq}{\begin{equation}}
\newcommand{\eeq}{\end{equation}}
\newcommand{\beqa}{\begin{eqnarray}}
\newcommand{\eeqa}{\end{eqnarray}}
\newcommand{\beg}{\begin{gather}}
\newcommand{\eeg}{\end{gather}}
\newcommand{\nol}{\nonumber}

\begin{document}

\title{How to use retarded Green's functions in de~Sitter spacetime}

\author{Atsushi HIGUCHI$^1$ and LEE Yen Cheong$^2$}

\affiliation{Department of Mathematics, University of York,
Heslington, York YO10 5DD, United Kingdom\\ email:
$^1$ah28@york.ac.uk, $^2$yl538@york.ac.uk}

\date{September 18, 2008}

\begin{abstract}
We demonstrate in examples that the covariant retarded Green's functions in electromagnetism and linearized gravity work as expected in de~Sitter spacetime.  We first clarify how retarded Green's functions should be used in spacetimes with spacelike past infinity such as de~Sitter spacetime.  In particular, we remind the reader of a general formula which gives the field for given initial data on a Cauchy surface and a given source (a charge or stress-energy tensor distribution) in its future.  We then apply this formula to three examples: (i) electromagnetism in the future of a Cauchy surface in Minkowski spacetime, (ii) electromagnetism in de~Sitter spacetime, and (iii) linearized gravity in de~Sitter spacetime.  In each example the field is reproduced correctly as predicted by the general argument.
In the third example we construct a linearized gravitational field from two equal point masses located at the ``North and South Poles" which is non-singular on the cosmological horizon and satisfies a covariant gauge condition and show that this field is reproduced by the retarded Green's function with corresponding gauge parameters.
\end{abstract}


\maketitle

\section{Introduction}

Over the past few decades quantum field theory in de~Sitter spacetime has been developed mainly because of its relevance to the inflationary universe scenario~\cite{Guth,Linde,Steinhardt,Sato1,Sato2,Kazanas} (see also Ref.~\cite{Starobinsky}).
In particular, infrared properties of linearized gravitational field have attracted much attention.
The mode functions for linearized gravity are similar to those of the minimally-coupled massless scalar field if one imposes gauge conditions natural to the spatially-flat coordinate system of de~Sitter spacetime~\cite{FordParker}. The two-point function of the minimally-coupled massless scalar field is known to be infrared (IR) divergent (see, e.g.\ Refs.~\cite{FordParker,Ratra,AllenInf,AllenFolacci}), and as a result one finds that the two-point function for linearized gravity is also IR divergent in this gauge.  However, since linearized gravity has gauge invariance, one needs to determine whether or not these IR divergences are a gauge artifact.  Indeed it has been shown that they can be gauged away (by gauge transformations which are not localized)~\cite{HiguchiNP,AllenFlat}.  It has also been shown that the two-point functions in gauges natural to other coordinate systems are IR finite~\cite{Hawkingetal,HiguchiWeeks}.

Infrared finite graviton Feynman propagators in the covariant gauge have also been constructed~\cite{AllenTuryn,Higuchi2}.  However, it has been claimed that the covariant Feynman propagators lead to incorrect retarded Green's functions which violate the linearized Einstein equations~\cite{Woodard,Miao}.  This assertion is partly based on the work of Antoniadis and Mottola~\cite{Antoniadis}, who claimed that in order for the scalar metric perturbation to obey the equations of motion, the de~Sitter invariant construction of the retarded Green's function must be abandoned (see also Ref.~\cite{AIT}).  However, it had been pointed out by Allen~\cite{AllenGravity} that the value of one of the gauge parameters used in their work introduces spurious IR divergences for the Feynman propagator in the sense that the IR divergences are absent for other values of gauge parameters, e.g.\ those adopted in Refs.~\cite{AllenTuryn,Higuchi2}.  As far as non-interacting linearized gravity is concerned, the conclusion of Antoniadis and Mottola is entirely due to these spurious divergences.  Moreover, by a closer inspection one finds that their retarded Green's function for the scalar metric perturbation, which is IR finite unlike the Feynman propagator, {\em does} satisfy the correct field equation contrary to their original claim~\cite{MottolaPrivate}.  Thus, any objection to the covariant retarded Green's function for linearized gravity based on their work is completely unfounded.

Another feature of the covariant retarded Green's functions that might appear to cast doubts on their validity is that, since they are causal, the field generated by an inertial point source in de~Sitter spacetime has support only in half of the spacetime~\cite{Woodard}. Thus, for example, the electric field generated by a freely falling point charge cannot satisfy the Gauss law because the flux out of a sufficiently large sphere around the charge vanishes, the field itself being zero there.  This apparent paradox can be resolved by recalling that in general the retarded Green's function is used to generate the field in the future of any Cauchy surface for given initial data on it and a source in its future. Thus, in order to reproduce the field using the retarded Green's function one needs to include the contribution from the initial data on spacelike past infinity in de~Sitter spacetime as well as that from the source: the field in de~Sitter spacetime is not determined by the source alone but is influenced also by the initial data on past infinity.  There may be philosophical uneasiness about the fact that the electric field on past infinity is necessarily nonzero if charged particles are present there~\cite{Philosophy}, and some authors have proposed that this field should be generated from the source by using half-advanced Green's function~\cite{Bicak}. (A related fact~\cite{Philosophy,Bicak} is that an electric charge does not generate a purely retarded electromagnetic field with support in its causal future.)  However, the retarded Green's functions, as a mathematical tool, are obviously not designed to generate the initial data on past infinity.

Thus, we find no compelling reason to doubt the validity of covariant retarded Green's functions either in electromagnetism or linearized gravity in de~Sitter spacetime.  However, in view of the recent claim that they do not work properly, it will be useful to demonstrate how they work in some examples.
The purpose of this paper is to show that the covariant retarded Green's functions for electromagnetism~\cite{AllenJacobson} and linearized gravity~\cite{Higuchi2} do indeed generate the fields which obey the equations of motion if we use the formula involving not only the source but also the initial data on a Cauchy surface.  Our calculations also serve as a check for the Feynman propagator obtained in Ref.~\cite{Higuchi2}. (See Ref.~\cite{Kouris} for a calculation of a physical quantity, the Weyl-tensor correlation function, using the two-point function of Ref.~\cite{Higuchi2}.)

The rest of the paper is organized as follows.
In Sec.~\ref{formalism} we address by using Carter-Penrose diagrams the necessity of including the contribution from past infinity as well as that from the source. Then,
we review the derivation of a general formula for reproducing the field from the initial data on a Cauchy surface and a source in its future in terms of the retarded Green's function.
We then apply this formula to three examples.
In order to illustrate its use in a simple setting, we first discuss how the electric field with a static charge is reproduced in the future of the $t=0$ Cauchy surface of Minkowski spacetime using the Cauchy data on the $t=0$ hypersurface in Sec.~\ref{case1}.
In Sec.~\ref{case2} we reproduce the electromagnetic field with two charges, one at the ``North Pole" and the other with the opposite sign at the ``South Pole".
We demonstrate that the general formula in Sec.~\ref{formalism} correctly reproduces the field throughout the whole manifold of de~Sitter spacetime.  In Sec.~\ref{case3} we first write down the linearized gravitational field with two equal mass points at the North and South Poles smooth at the horizon and satisfying a covariant gauge condition.
Then we show that the linearized gravitational field on the entire spacetime is recovered using the general formula in Sec.~II as expected.
Throughout this paper we call the contribution to the field coming from the source the \textit{source field} and the contribution coming from the initial data the \textit{initial field} and use natural Planck units $\hbar=c=G=1$ and the sign convention $-+++$.

\section{General formula for the use of retarded Green's function} \label{formalism}

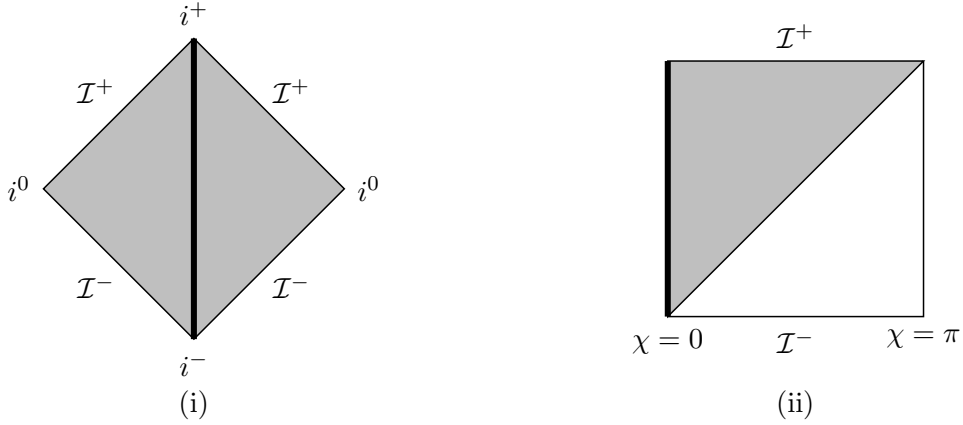
\begin{figure}[ht]
\begin{center}
\begin{pspicture}(-6,-0.5)(6,4.5)
\pspolygon[linewidth=0.2mm, fillstyle=solid, fillcolor=lightgray](-6,2)(-4,4)(-2,2)(-4,0)(-6,2)
\pspolygon[linewidth=0.2mm, fillstyle=solid, fillcolor=lightgray](2.3,0.3)(5.7,3.7)(2.3,3.7)(2.3,0.3)
\psline[linewidth=0.2mm](2.3,0.3)(5.7,0.3)(5.7,3.7)
\psline[linewidth=0.8mm](-4,0)(-4,4)
\psline[linewidth=0.8mm](2.3,0.3)(2.3,3.7)
    \uput[u](-4,4){$i^+$}
    \uput[d](-4,0){$i^-$}
    \uput[r](-2,2){$i^0$}
    \uput[l](-6,2){$i^0$}
    \rput(-2.7,3.3){$\mathcal{I}^+$}
    \rput(-5.3,3.3){$\mathcal{I}^+$}
    \rput(-5.3,0.7){$\mathcal{I}^-$}
    \rput(-2.7,0.7){$\mathcal{I}^-$}
    \uput[u](4,3.7){$\mathcal{I}^+$}
    \uput[d](4,0.3){$\mathcal{I}^-$}
    \uput[d](2.3,0.3){$\chi=0$}
    \uput[d](5.7,0.3){$\chi=\pi$}
    \uput[d](-4,-0.5){(i)}
    \uput[d](4,-0.5){(ii)}
\end{pspicture}
\end{center}
\caption{(i) The Carter-Penrose diagram of Minkowski spacetime;
(ii) Carter-Penrose diagram of de~Sitter spacetime [see Eq.~(\ref{global}) for the definition of $\chi$].
The field coming from a classical source (its world line represented by a vertical bold line) is nonzero in the shaded region.} \label{Penrose}
\end{figure}

The electromagnetic and gravitational fields generated by the retarded Green's function propagate at most at the speed of light.  On the other hand the Gauss law in electromagnetism and a similar law in linearized gravity in spacetime with a Killing vector field imply that
the total outward flux of the field out of an arbitrary closed spatial surface is equal to the total conserved charge enclosed by such a surface. These two facts would appear to invalidate the use of covariant retarded Green's functions in de~Sitter spacetime~\cite{Woodard}.
This apparent paradox can be best understood using Carter-Penrose diagrams.
The Carter-Penrose diagrams for Minkowski and de~Sitter spacetimes are shown in Fig.~\ref{Penrose} (see, e.g.\ Ref.~\cite{HawkingEllis}).
The bold line represents the world line of an inertial source originating at past infinity $i^-$ ($\mathcal{I}^-$) and ending at future infinity $i^+$ ($\mathcal{I}^+$) in Minkowski (de~Sitter) spacetime.
The field causally generated by the source along its world line has support in the shaded region.
For electromagnetism, for example, the field generated using the retarded Green's function by the charge cannot satisfy the Gauss law on any constant-time hypersurface in de~Sitter spacetime since the electromagnetic field vanishes outside a sphere of some radius at each time, as can easily be seen by drawing a horizontal line across the Carter-Penrose diagram, whereas in Minkowski spacetime the Gauss law is satisfied by the field generated by the retarded Green's function from the source.
The cause of this apparent difficulty in de~Sitter spacetime is the spacelike nature of past infinity $\mathcal{I}^-$.

In fact there is nothing wrong with the retarded Green's functions de~Sitter spacetime.  In some sense this apparent difficulty is caused because one is implicitly requiring too much of the retarded Green's function.  To understand this point one needs to recall how the retarded Green's function is used to generate the field in the future of an arbitrary Cauchy surface.
There is a formula, which is the key formula of this paper, for generating the field in terms of the retarded Green's function for a given initial data on a Cauchy surface and a given source.  This formula is well known for scalar fields (see, e.g.\ Ref.~\cite{Wald}) and is known for other bosonic fields as well, but it will be useful to review its derivation here.

Suppose the Lagrangian density for a local field $A_I$
(with an upper case Latin index representing a set of indices) is given by
\beq
    \mathcal{L} = \frac{\sqrt{-g}}{2}\[T^{aIbJ}\nabla_aA_I\nabla_bA_J + S^{IJ}A_IA_J\]\,,
\eeq
where $T^{aIbJ}$ and $S^{IJ}$ are tensors satisfying $T^{aIbJ}=T^{bJaI}$ and $S^{IJ}=S^{JI}$ and independent of $A_I$.
The conjugate momentum current is defined by
\beq
    \pi^{cJ} = \frac{1}{\sqrt{-g}}\frac{\partial\mathcal{L}}{\partial(\nabla_cA_J)}
        = T^{cJaI}\nabla_aA_I \,. \label{conj_mom_curr}
\eeq
Let $L_\pi$ be the differential operator defined by the equation
\beq
    (L_\pi A)^{cI} = \pi^{cI} \,. \label{L pi}
\eeq
Thus, the operator $L_{\pi}$ maps the field $A_{I}$ to its conjugate momentum
current $\pi^{cI}$.
The Euler-Lagrange equations read
\beq
    L^{IJ}A_{J} \equiv \nabla_c \pi^{cI} - S^{IJ}A_{J} = 0 \,. \label{EL equation}
\eeq
We assume that there are unique retarded and advanced Green functions, $G^R_{II'}(x,x')$ and $G^A_{II'}(x,x')$, satisfying
\beq
    L_x^{IJ}G^{R/A}_{JI'}(x,x') = {\delta^{I}}_{I'}\delta^{4}(x,x') \,, \label{DefineGreen_R}
\eeq
where the (un)primed indices refer to the (un)primed point ($x$) $x'$ and where the differential operator $L_x^{IJ}$ acts at point $x$.  The retarded (advanced) Green's function is also required to vanish if point $x$ is not in the causal future (past) of point $x'$, i.e.\ if there is no future-directed (past-directed) causal curve from $x'$ to $x$.  The delta function ${\delta^{I}}_{I'}\delta^4(x,x')$ is defined by the property that
\beq
\int d^4x'\sqrt{-g'}{\delta^I}_{I'}\delta^4(x,x')A^{I'}(x') = A^I(x)  \label{AHdefdelta}
\eeq
if $A^I(x)$ is smooth and compactly supported.
It can be shown, as is well known, that
\beq
L^{I'J'}_{x'}G^R_{IJ'}(x,x') = {\delta^{I'}}_I\delta^4(x',x)\,, \label{AHGreen}
\eeq
which implies $G^A_{II'}(x,x') = G^R_{I'I}(x',x)$ by the assumed uniqueness of the advanced Green's function.
We present a proof of this formula in Appendix~\ref{AppendixA}.

Let us denote by $D^+(\Sigma)$ the future domain of dependence of the Cauchy surface $\Sigma$.
(See, e.g.\ Refs.~\cite{HawkingEllis,WaldTextbook} for the definition of the domain of dependence.)
Let $x$ be a point in the future of $\Sigma$, i.e.\ $x\in D^+(\Sigma)$,
and $x'$ be a point on the Cauchy surface $\Sigma$, i.e.\ $x'\in\Sigma$.
Consider a source represented by the current $J^I$ that is nonzero in the region $D^+(\Sigma)$ .
The unique solution to the inhomogeneous equations of motion
\beq
L^{IK}A_K = \nabla_c \pi^{cI} - S^{IK}A_{K} = J^I  \label{AHinhomo}
\eeq
in the future of a Cauchy surface $\Sigma$ with given initial data on $\Sigma$ can be expressed in terms of the retarded Green's function $G^R_{II'}(x,x')$.
Let $(A_{I'},\pi^{c'I'}N_{c'})$ be the initial data on $\Sigma$, where $N_{c'}$ is the past-directed unit normal on $\Sigma$. (Note that the time component of $N_{c'}$ is positive.)  Let $\Sigma'$ be another Cauchy surface in the future of $\Sigma$ and let $x \in U\equiv D^+(\Sigma)\cap D^-(\Sigma')$, where $D^-(\Sigma')$ is the past domain of dependence of $\Sigma'$. Then
\beqa
    A_I(x) &=& \int_U d^4x'\sqrt{-g(x')}\delta^{4}(x,x'){\delta_I}^{I'}A_{I'}(x') \nol \\
            &=& \int_U d^4x'\sqrt{-g(x')}L^{I'J'}_{x'}G^R_{IJ'}(x,x')A_{I'}(x') \,,
            \label{From_initial}
\eeqa
where we have used Eq.~(\ref{AHGreen}). 
Using Eq.~(\ref{AHinhomo}), we find that Eq.~(\ref{From_initial}) can be expressed as
\beqa
    A_I(x)
        &=& \int_U d^4x'\sqrt{-g(x')}\left\{ \nabla_{c'} \[ {(L_{\pi}G^R)_I}^{c'I'}(x,x')A_{I'}(x')
                 - G^R_{IJ'}(x,x')\pi^{c'J'}(x') \] \right.\nol \\
            &&     \qquad +G^R_{II'}(x,x')J^{I'}(x')\nol\\
            && \qquad - \left. \[ {(L_{\pi}G^R)_I}^{c'I'}(x,x')\nabla_{c'}A_{I'}(x')
            - \nabla_{c'}\(G^R_{IJ'}(x,x')\)\pi^{c'J'}(x') \]\right\} \,. \label{AHadded}
\eeqa
By Eq.~(\ref{conj_mom_curr}) and the equation ${(L_{\pi}G^R)_I}^{c'I'}(x,x') = T^{c'I'a'J'}\nabla_{a'}G^R_{IJ'}(x,x')$
together with the symmetry property of $T^{c'I'a'J'}$,
the last two terms on the right-hand side of Eq.~(\ref{AHadded}) cancel out.
Then, by the generalized Gauss theorem we find
\beqa
    A_I(x) &=& \int_U d^4x'\sqrt{-g'}\,G^R_{II'}(x,x')J^{I'}(x')\nol\\
    && +\int_{\Sigma'} d\Sigma_{c'} \[ {(L_{\pi}G^R)_I}^{c'I'}(x,x')A_{I'}(x')
                - G^R_{IJ'}(x,x')\pi^{c'J'}(x') \] \nol \\
        && - \int_{\Sigma} d\Sigma_{c'} \[ {(L_{\pi}G^R)_I}^{c'I'}(x,x')A_{I'}(x')
                - G^R_{IJ'}(x,x')\pi^{c'J'}(x') \] \,, \label{AHthiseq}
\eeqa
where $d\Sigma_{c'}=d\Sigma N_{c'}$. The second term, the integral over $\Sigma'$, vanishes because $G^R_{II'}(x,x')=0$ if $x$ is not in the causal future of $x'$. Hence,
\beq
A_I(x) = A_I^{(S)}(x) + A_I^{(I)}(x)\,,  \label{key formula}
\eeq
where the {\em source field} $A_I^{(S)}(x)$ and the {\em initial field} $A_I^{(I)}(x)$ are given by
\beqa
    A^{(S)}_{I}(x) &=& \int_{D^+(\Sigma)}d^4x'\sqrt{-g(x')} G^R_{II'}(x,x')J^{I'}(x') \,, \label{From_source}\\
    A_I^{(I)}(x) &=& \int_{\Sigma} d\Sigma_{c'} \left[\pi^{c'I'}(x')G^R_{II'}(x,x')
        - A_{I'}(x'){(L_\pi G^R)_I}^{c'I'}(x,x') \right]\,. \qquad \label{key formula_original}
\eeqa
We have changed the integration domain for $A^{(S)}_I(x)$ from $U=D^{+}(\Sigma)\cap D^{-}(\Sigma')$ to $D^{+}(\Sigma)$ because $G^R_{II'}(x,x')=0$ unless $x'\in D^{-}(\Sigma')$ by the assumption that $x \in U$.
The field $A_I(x)$ given by Eq.~(\ref{key formula})
is the unique solution to the inhomogeneous equations (\ref{AHinhomo}) for the given initial data on $\Sigma$.

In the next section we show how this formula can be used to reproduce the electric field in the future of the $t=0$ Cauchy surface in Minkowski spacetime with a static point charge.  In Secs.~\ref{case2} and \ref{case3} we show in examples that the retarded Green's functions for electromagnetism~\cite{AllenJacobson} and linearized gravity~\cite{Higuchi2} in de~Sitter spacetime generate the correct fields through this formula.

\section{Electromagnetic field in Minkowski spacetime} \label{case1}

\begin{figure}[ht]
\begin{center}
\begin{pspicture}(-8,-1)(2,4.5)
    \psline[linewidth=0.2mm]{<->}(-8,4)(-8,0)(-4,0)
        \uput[r](-4,0){$|\mathbf{x}|$}
        \uput[u](-8,4){$A^t$}
        \pscurve[linewidth=0.2mm, linestyle=dotted](-7.35,1.5)(-7,1)(-6.5,0.7)(-4.5,0.3)
        \pscurve[linewidth=0.2mm](-7.7,3.5)(-7.5,2)(-7.35,1.5)
        \psline[linewidth=0.2mm](-7.35,1.5)(-7.35,0)
            \psline[linewidth=0.2mm]{<-}(-7.35,-0.1)(-7.35,-0.6)
                \uput[d](-7.5,-0.6){$|\mathbf{x}|=t$}
    \psline[linewidth=0.2mm]{->}(-3.5,2)(-2.5,2)
        \uput[u](-3,2){Time}
    \psline[linewidth=0.2mm]{<->}(-2,4)(-2,0)(2,0)
        \uput[r](2,0){$|\mathbf{x}|$}
        \uput[u](-2,4){$A^t$}
        \pscurve[linewidth=0.2mm](-1.7,3.5)(-1,1)(0.5,0.4)
        \psline[linewidth=0.2mm, linestyle=dotted](0.5,0.4)(1.5,0.3)
        \psline[linewidth=0.2mm](0.5,0.4)(0.5,0)
            \psline[linewidth=0.2mm]{<-}(0.5,-0.1)(0.5,-0.6)
                \uput[d](0.5,-0.6){$|\mathbf{x}|=t$}
\end{pspicture}
\end{center}
\renewcommand{\figurename}{Fig.}
\caption{The spreading of the source field as the time elapses.}
\label{Evol_A_source}
\end{figure}
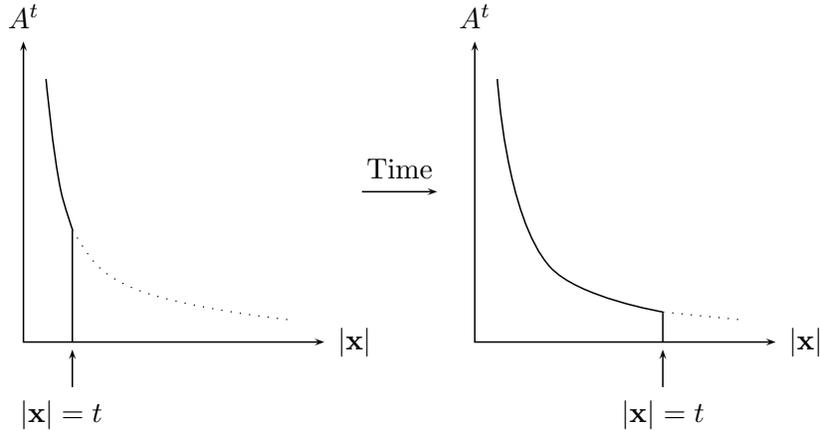

In this section we demonstrate how the formula derived in the previous section works in the simple example of the static electric field compatible with a static electric charge in the future half of Minkowski spacetime.
The Lagrangian density for the massless spin-1 vector field $A^a$ in the covariant gauge is given by
\beq
    \mathcal{L} = -\frac{1}{4}F_{ab}F^{ab} - \frac{1}{2\zeta}\(\partial_aA^a\)^2\,, \label{AHFeynman}
\eeq
where the electromagnetic field tensor is $F_{ab}=\partial_aA_b-\partial_bA_a$.  We adopt the Feynman gauge $\zeta = 1$.  As is well known, the retarded Green's function in this gauge is
\beq
    {G^a}_{a'}(x,x') = \frac{\delta^a_{a'}\delta(t-t'-|\mathbf{x-x'}|)}{4\pi|\mathbf{x-x'}|}\theta(t-t')\,.
    \label{Green_flat}
\eeq
\begin{figure}[ht]
\begin{center}
\begin{pspicture}(-3,-0.5)(3,3.5)
 \pspolygon[linestyle=none, fillstyle=solid, fillcolor=lightgray](0,0)(-2,2)(2,2)(0,0)
 \psline[linewidth=0.8mm](0,-0.5)(0,2.5)
 \psline[linewidth=0.2mm]{->}(0,2.5)(0,3)
    \uput[u](0,3){$t$}
 \psline[linewidth=0.2mm]{<-}(2.2,0)(2.6,0)
    \uput[r](2.6,0){$t=0$}
 \psline[linewidth=0.5mm, linestyle=dotted](-2,0)(2,0)
    \uput[d](-2,0){Initial surface $\Sigma$}
\end{pspicture}
\end{center}
\caption{A static charge $q$ at the center producing the field in the shaded region} \label{Flat_case}
\end{figure}
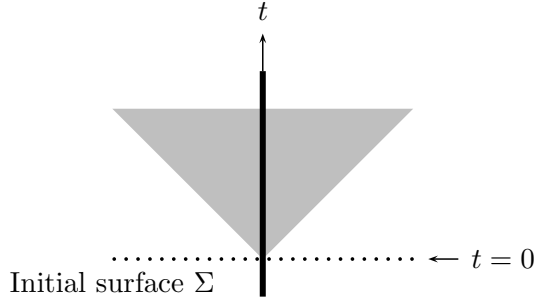
In Fig.~\ref{Flat_case} we show the support of the source field from a static point charge if the initial surface is taken to be the $t=0$ hypersurface.
Now, let a static point charge $q$ be placed at the spatial origin.
Then the corresponding 4-current is $J^a(x)=(q\delta^3(\mathbf{x}),\mathbf{0})$.
With the retarded Green's function in Minkowski spacetime given by Eq.~(\ref{Green_flat})
we find the source field to be
\beqa
    A^{a(S)}(x) &=& \int_{t'>0} d^4x'
    \frac{q}{4\pi}\frac{\delta(t-t'-|\mathbf{x-x'}|)}{|\mathbf{x-x'}|}\theta(t-t')\delta^a_t\delta^3(\mathbf{x'}) \nol \\
        &=& \frac{q}{4\pi|\mathbf{x}|}\theta(t-|\mathbf{x}|)\delta^a_t \,, \label{AHsource}
\eeqa
where $\theta(t-t')$ is the Heaviside step function.
We note that the Heaviside step function appears naturally due to the causal nature of the retarded Green's function.
As illustrated in Fig.~\ref{Evol_A_source}, the source field spreads as the time elapses.

Next, we turn our attention to the initial field.
We use the input field $A^{a'}=(A^{t'},\mathbf{0})$ with
\beq
    A^{t'}(\mathbf{x'},t'=0) = \frac{q}{4\pi|\mathbf{x'}|} \,, \label{AHfield}
\eeq
where $\mathbf{x'}$ is the position vector on initial surface.
This field is generated on the initial Cauchy surface by the charge before the time $t'=0$.
In the Feynman gauge $\zeta=1$ we find that the Lagrangian density can be written as
\beq
    \mathcal{L} = \tfrac{1}{2}T^{abcd}\partial_aA_b\cdot\partial_cA_d\,,
\eeq
where
\beq
T^{abcd}\equiv g^{bc}g^{ad}-g^{ac}g^{bd}-g^{ab}g^{cd}\,.
\eeq
Hence the conjugate momentum current is
\beq
    \pi^{ca} = T^{cadb}\partial_dA_b = -F^{ca}\,, \label{pi1}
\eeq
where we have used the fact that the field (\ref{AHfield}) satisfies the Lorenz condition $\partial_a A^a = 0$.
The only nonzero component of the past-directed unit normal to the hypersurface $\Sigma$ is $N_{t'}=1$.
We find from Eq.~(\ref{key formula_original}) that the initial field is given by
\beq
    A^{a(I)}(x)
        = \int_{t'=0} d\Sigma_{t'} \[ \partial^{a'}A^{t'}\cdot {G^a}_{a'} - A_{a'}\( -\partial^{t'}G^{aa'}
        + \partial^{a'}G^{at'} - g^{t'a'}\partial_{b'}G^{ab'} \)\]\,. \label{AHfirst}
\eeq
Noting that $A_{a'}\( -\partial^{t'}G^{aa'} + \partial^{a'}G^{at'}\) = 0$ and dropping the total spatial divergence $\sum_{i'=1}^3\partial_{i'}(A^{t'}G^{ai'})$, we find
\beqa
	A^{a(I)}(x)
        &=& -\int_{t'=0} d^3x' A^{t'}\frac{\partial}{\partial t'}{G^a}_{t'} \nol \\
        &=& \int d^3x' \frac{q}{4\pi|\mathbf{x'}|}
        \frac{\delta'(t-|\mathbf{x-x'}|)}{4\pi|\mathbf{x-x'}|} \delta^a_t \nol \\
        &=& \frac{q}{4\pi|\mathbf{x}|}\theta(|\mathbf{x}|-t)\delta^a_t \,. \label{AHinitial}
\eeqa
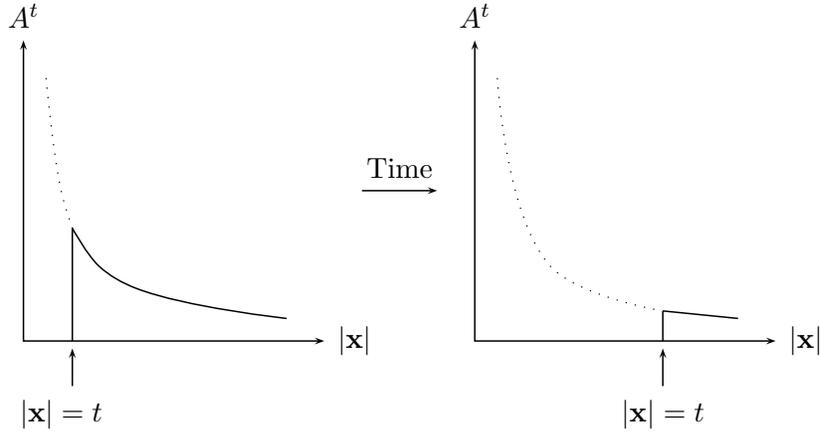
\begin{figure}[ht]
\begin{center}
\begin{pspicture}(-8,-1)(2,4.5)
    \psline[linewidth=0.2mm]{<->}(-8,4)(-8,0)(-4,0)
        \uput[r](-4,0){$|\mathbf{x}|$}
        \uput[u](-8,4){$A^t$}
        \pscurve[linewidth=0.2mm](-7.35,1.5)(-7,1)(-6.5,0.7)(-4.5,0.3)
        \pscurve[linewidth=0.2mm, linestyle=dotted](-7.7,3.5)(-7.5,2)(-7.35,1.5)
        \psline[linewidth=0.2mm](-7.35,1.5)(-7.35,0)
            \psline[linewidth=0.2mm]{<-}(-7.35,-0.1)(-7.35,-0.6)
                \uput[d](-7.5,-0.6){$|\mathbf{x}|=t$}
    \psline[linewidth=0.2mm]{->}(-3.5,2)(-2.5,2)
        \uput[u](-3,2){Time}
    \psline[linewidth=0.2mm]{<->}(-2,4)(-2,0)(2,0)
        \uput[r](2,0){$|\mathbf{x}|$}
        \uput[u](-2,4){$A^t$}
        \pscurve[linewidth=0.2mm, linestyle=dotted](-1.7,3.5)(-1,1)(0.5,0.4)
        \psline[linewidth=0.2mm](0.5,0.4)(1.5,0.3)
        \psline[linewidth=0.2mm](0.5,0.4)(0.5,0)
            \psline[linewidth=0.2mm]{<-}(0.5,-0.1)(0.5,-0.6)
                \uput[d](0.5,-0.6){$|\mathbf{x}|=t$}
\end{pspicture}
\end{center}
\renewcommand{\figurename}{Fig.}
\caption{The behavior of the initial field as the time elapses.}
\label{Evol_A_initial}
\end{figure}
\newline
\noindent
Fig.~\ref{Evol_A_initial} illustrates that the initial field eventually becomes zero for any given point as time elapses.
{}From Eqs.~(\ref{AHsource}) and (\ref{AHinitial}) we find
\beqa
    A^a(x) & = & A^{a(S)}(x)+A^{a(I)}(x) \nol \\
    & = & \frac{q}{4\pi|\mathbf{x}|}\delta^a_t \,,
\eeqa
recovering the correct field.
The calculations in de~Sitter spacetime are much more complicated, but we shall see that the initial field fills up the region that is not in the causal future of the source in a way similar to this example.

\section{Electromagnetic field in de~Sitter spacetime} \label{case2}

The metric of de~Sitter spacetime can be found by regarding this spacetime as the hypersurface $-(X^0)^2 + \sum_{i=1}^4 (X^i)^2 = H^{-2}$ in $5$-dimensional Minkowski spacetime with the metric
$
ds_M^2 = -(dX^0)^2 + \sum_{i=1}^4 (dX^i)^2$.
This hypersurface can be parametrized as
$X^0 = H^{-1}\tan \tau$,
$X^1  =  H^{-1}\sec \tau \cos \chi$,
$X^2  =  H^{-1}\sec \tau \sin\chi \cos\theta$,
$X^3  =  H^{-1}\sec \tau \sin\chi \sin\theta \cos\phi$,
$X^4  =  H^{-1}\sec \tau \sin\chi\sin\theta \sin\phi$, with the range of the parameters given by $\tau \in (-\pi/2,\pi/2)$, $\chi, \theta \in [0,\pi]$ and $\phi \in [0,2\pi)$.  The resulting metric, which is valid for the whole spacetime, is
\beq
ds^2 = \frac{1}{H^2\cos^2 \tau}\left( -d\tau^2 + d\chi^2 + \sin^2\chi d\Omega^2\right)\,, \label{global}
\eeq
where
\beq
d\Omega^2 = d\theta^2 + \sin^2\theta\,d\phi^2\,.
\eeq
In this section we reproduce the electromagnetic field compatible with a charge $q$ at $\chi=0$ (the ``North Pole") and another charge $-q$ at $\chi=\pi$ (the ``South Pole") using the retarded Green's function. [As is well known, the total charge on a compact space must vanish (see, e.g.\ Ref.~\cite{BrillDeser}).  Hence it is not possible to have just one point charge in de~Sitter spacetime.]
The massless spin-1 Feynman propagator $Q_{aa'}(x,x')$ in the covariant Feynman gauge in de~Sitter spacetime background, satisfying
\beq
(-\nabla_b \nabla^b +3H^2)Q_{aa'}(x,x') = -i g_{aa'}\delta^{4}(x,x')\,,
\eeq
where $H$ is the Hubble constant of de~Sitter spacetime,
has been calculated by Allen and Jacobson~\cite{AllenJacobson}.
We need to give some definitions in Ref.~\cite{AllenJacobson} in order to present their result.  Given a pair of spacelike separated points $x$ and $x'$ we define $\mu(x,x')$ to be the geodesic distance between them.  The variable $z$ is defined by
\beq
    z = z(x,x') = \frac{1}{2}(1+\cos H\mu) \,. \label{z}
\eeq
We also define the unit vectors $n_a \equiv \nabla_a\mu$ at $x$ and $n_{a'} \equiv \nabla_{a'}\mu$ at $x'$.  These are tangent to the geodesic between $x$ and $x'$.  The vector $n_a$ is given by
\begin{equation}
    n_a = -\frac{1}{2H\sqrt{z(1-z)}}\partial_a \cos H\mu \,. \label{n}
\end{equation}
We also define the parallel propagator, $g_{aa'}$, as follows: given a vector $V^a$ at $x$, the vector $W^{a'}$ obtained by parallelly transporting $V^a$ to $x'$ along the geodesic is given by $W^{a'}={g_a}^{a'}V^a$.  It can readily be seen that $V^a = {g^a}_{a'}W^{a'}$.  The parallel propagator is given explicitly as
\begin{equation}
    g_{aa'} = \frac{1}{H^2}\left( \partial_a\partial_{a'}\cos H\mu
                - \frac{1}{2z}\partial_a\cos H\mu\cdot\partial_{a'}\cos H\mu \right) \,. \label{prllprop}
\end{equation}
Then, the Feynman propagator of Allen and Jacobson is
\beq
    Q_{aa'}(x,x') = \alpha(z)g_{aa'}(x,x') + \beta(z)n_a(x) n_{a'}(x')\,,
\eeq
where
\begin{eqnarray}
    \alpha(z) & = & \frac{H^2}{48\pi^2}\left[\frac{3}{1-z} + \frac{1}{z} + \left(\frac{2}{z} + \frac{1}{z^2}\right)
                    \log(1-z)\right]\,, \label{alpha from Q}\\
    \beta(z) & = & \frac{H^2}{24\pi^2}\left[1-\frac{1}{z} + \left(\frac{1}{z}-\frac{1}{z^2}\right)\log(1-z)\right] \,. \label{beta from Q}
\end{eqnarray}
The Feynman propagator $Q_{aa'}(x,x')$ for points that are not spacelike separated is defined by analytic continuation.
It is analytically continued around the singularity at $\mu^2=0$, i.e.\ $z=1$, by replacing $\mu^2$ by $\mu^2 + i\epsilon$, where $\epsilon$ is an ``infinitesimal" positive number.  As a result $1-z$ is replaced by $1-z + i\epsilon$.

The retarded Green's function is proportional to the difference of the values of the two-point function across the branch cut $[1,\infty)$ on the real axis. (The relation between various Green's functions can be found, e.g.\ in Ref.~\cite{Birrell} for scalar fields.  The Green's functions for bosonic higher-spin fields are related in the same way.)  We note that
the singular terms at $z=0$ in $\beta(z)$ cancel out.  The retarded Green's function $G_{aa'}(x,x')$ is given by
\beq
    G_{aa'}(x,x') = -i\theta(\tau-\tau')\left[Q_{aa'}(z+i\epsilon) - Q_{aa'}(z-i\epsilon)\right]\,,
\eeq
where $\tau$ and $\tau'$ are the time coordinates of points $x$ and $x'$, respectively.
The explicit form of $G_{aa'}(x,x')$ can be found by using the Sokhotski formula,
\beq
    \frac{1}{1-z \pm i\epsilon} = {\rm P}\,\frac{1}{1-z} \mp i\pi \delta(1-z)\,, \label{Sokhotski1}
\eeq
%
where P denotes the principal part.
We also use
\beq
    \log(1-z \pm i\epsilon) = \log|1-z| \pm i\pi\theta(z-1)\,. \label{Sokhotski4}
\eeq
The resulting retarded Green's function is
\beq
    G_{aa'}(x,x')
                    = G^{(\delta)}_{aa'}(x,x')\delta(1-z) + G^{(\theta)}_{aa'}(x,x')\theta(z-1)\,, \label{AHGret}
\eeq
where
\beqa
    G^{(\delta)}_{aa'}(x,x') &=& \frac{H^2}{8\pi}g_{aa'}(x,x') \,, \label{G_delta}\\
    G^{(\theta)}_{aa'}(x,x') &=& -\frac{H^2}{12\pi}\left[\left(\frac{1}{z}+\frac{1}{2z^2}\right)g_{aa'}(x,x')
                                + \frac{z-1}{z^2}n_a(x)n_{a'}(x')\right] \,. \label{G_theta}
\eeqa
Here, the factor $\theta(\tau-\tau')$ is understood though not explicitly written since it will not affect our calculations.
Note that the singularity of $n_an_{a'}$ as $z\to 1$ [see Eq.~(\ref{n})] is canceled by the factor $z-1$.

A solution to Maxwell's equations with the two charges at the Poles is found most readily in the static coordinate system with the metric
\beq
    ds^2 = -(1-H^2R^2)dT^2 + (1-H^2R^2)^{-1}dR^2 + R^2d\Omega^2\,. \label{AHstatmet}
\eeq
This coordinate system is related to the global coordinate system (\ref{global}) as
\beqa
T & = & \frac{1}{2H}\log\frac{\cos\chi + \sin \tau}{\cos\chi - \sin \tau}\,,\\
R & = & \frac{\sin \chi}{H\cos \tau}\,.
\eeqa
The static electric field with charge $q$ at the origin $R=0$ is given by
\beq
    F^{TR} = \frac{q}{4 \pi R^2} \,.  \label{AHEM}
\eeq
This equation is solved by
\beq
    A_T = - \frac{q}{4\pi}\frac{1-HR}{R} \,, \label{A_T}
\eeq
with all other components vanishing.
We have added a constant $qH/4\pi$ to a more natural solution $A_T = -q/4\pi R$ to make the scalar quantity $A_a A^a$ non-singular at the horizon $R=1/H$.
This field satisfies the field equation $(\Box - 3H^2)A_a = 0$ at $R\neq 0$ and the Lorenz gauge condition $\nabla_a A^a = 0$. Hence it should be reproduced by the retarded Green's function (\ref{AHGret}).
In the global coordinate system this field is
\beqa
A_\tau & = & - \frac{q}{4\pi}\frac{\cos \tau \cot\chi}{\cos \tau + \sin \chi}\,,\\
A_\chi & = & - \frac{q}{4\pi}\frac{\sin \tau}{\cos \tau + \sin \chi}\,.
\eeqa
Under the transformation $\chi \mapsto \pi - \chi$, which interchanges the North and South Poles, the field $A_a$ simply changes its sign.  Thus, it is clear that there are two charges, $q$ at the North Pole and $-q$ at the South Pole.

Our calculations to reproduce the field (\ref{A_T}) by the retarded Green's function
can be performed most simply in the conformally-flat coordinate system with the metric
\beq
    ds^2 = \frac{1}{H^2\lambda^2} (-d\lambda^2 + dr^2 + r^2\Omega^2)\,.
\eeq
This coordinate system covers only the upper half of de~Sitter spacetime, which is the causal future of the charge at the North Pole, represented by the shaded region of the Carter-Penrose diagram in Fig.~\ref{Penrose}.  This coordinate system is related to the static coordinate system by
\beqa
T & = & - \frac{1}{2H}\log\left[H^2 (\lambda^2 - r^2)\right]\,,\\
R & = & \frac{r}{H\lambda}\,.
\eeqa
Note that the time variable $\lambda$ increases toward the past.  The nonzero components of the electromagnetic potential in the new coordinate system are
\beqa
A_\lambda & = & \frac{q}{4\pi}\left( \frac{1}{r} - \frac{1}{r+\lambda}\right)\,,\label{AHnewcoord1}\\
A_r & = & - \frac{q}{4\pi}\frac{1}{r+\lambda}\,. \label{AHnewcoord2}
\eeqa
We need to express $\cos H\mu(x,x')$ in terms of the coordinates of the two points $x$ and $x'$ in order to find the explicit expressions of $z$, $n_a$ and $g_{aa'}$ through Eqs.~(\ref{z}), (\ref{n}) and (\ref{prllprop}).
By writing down $H^{-2}\cos H \mu = -X^{0}X^{0'} + \sum_{i=1}^4 X^i X^{i'}$ in conformally-flat coordinates we find
\begin{equation}
    \cos H\mu = \frac{\lambda^2 + \lambda'^2 - r^2 - r'^2
                + 2rr'[\cos\theta\cos\theta'+\sin\theta\sin\theta'\cos(\phi-\phi')]}
                {2\lambda\lambda'} \,. \label{cosmu}
\end{equation}
The function $\cos H\mu$ is extended naturally to the case where the two points are not spacelike separated
by adopting this formula as it is.

Now, we calculate the source field from the charge $q$ at the North Pole.
We adopt the convention that the unprimed coordinates are those of the point where the field is calculated whereas the primed coordinates are those of the charge.
The $4$-current corresponding to the charge at the North Pole can readily be determined by noting that it must be a conserved current.
It is given by $J^{a'} = (J^{\lambda'},\mathbf{0})$, where
\beq
    J^{\lambda'}=-qH^4\lambda'^4\delta^3(\mathbf{x'}).  \label{AH4current}
\eeq
Here the delta function is defined to satisfy
\beq
	\int d^3x'\sqrt{\gamma'} \, \delta^3(\mathbf{x}') = 1 \,,
\eeq
where $d^3x'\sqrt{\gamma'}=dr'd\theta'd\phi'\,r^{\prime 2}\sin\theta'$.  We have from Eqs.~(\ref{cosmu}) and (\ref{AH4current})
\beqa
\sqrt{-g'}\delta(1-z)J^{\lambda'} & = & -\frac{2q\lambda\lambda'}{r}\delta(\lambda'-\lambda - r)\sqrt{\gamma'}\delta^3(\mathbf{x}')\,,\label{AHdelta}\\
\sqrt{-g'}\theta(1-z)J^{\lambda'} & = & -q\theta(\lambda'-\lambda -r)\sqrt{\gamma'}\delta^3(\mathbf{x}')\,. \label{AHtheta}
\eeqa
We first cut off the current at $\lambda'=\lambda_0$ by using the $4$-current $J^{a'}\theta(\lambda_0 - \lambda')$ and then take the limit $\lambda_0\to \infty$ in the end. (Recall that $\lambda'$ increases toward the past.)
Substituting the retarded Green's function (\ref{AHGret}) in Eq.~(\ref{From_source}) and using Eqs.~(\ref{AHdelta}) and (\ref{AHtheta}), we find
\beq
	A^{(S)}_a = \left.-q\left(\left.\frac{2\lambda\lambda'}{r}G^{(\delta)}_{a\lambda'}\right|_{\lambda'=\lambda+r}
			+ \int_{\lambda+r}^{\infty} d\lambda' \, G^{(\theta)}_{a\lambda'} \right)
            \right|_{\mathbf{x'}=0}\theta(\lambda_0-\lambda-r)\,,
\eeq
and hence
\begin{eqnarray}
	A^{(S)}_{\lambda}
&=& \frac{q}{4\pi} \left( \frac{1}{r} - \frac{1}{r+\lambda} + \frac{1}{3\lambda} \right)
                    \theta(\lambda_0-\lambda-r) \,, \label{Source:At} \\
    A^{(S)}_r &=& - \frac{q}{4\pi(\lambda+r)} \theta(\lambda_0-\lambda-r) \,, \label{Source:Ar}
\end{eqnarray}
with $A_{\theta}^{(S)}=A_{\phi}^{(S)}=0$.  The source field $A_a^{(S)}$ vanishes for $r > \lambda_0-\lambda$, i.e.\ outside the causal future of the world line of the truncated charge $J^{a'}\theta(\lambda_0-\lambda')$, but in the limit $\lambda_0\to \infty$ the field becomes nonzero everywhere in the causal future of the charge at the North Pole, which coincides with the part of de~Sitter spacetime covered by this conformally-flat coordinate system.  Thus, the Heaviside function can be dropped in this limit.
We note that for $r < \lambda_0 - \lambda$ the field $A^{(S)}_a$ does not agree with that given by Eqs.~(\ref{AHnewcoord1}) and (\ref{AHnewcoord2}) nor does it satisfy the Lorenz gauge condition because of the term $1/3\lambda$ in Eq.~(\ref{Source:At}).  Nevertheless, the source field gives the correct field strength $F_{ab}$ because the extra term is of the form
$\nabla_a \Lambda$ with $\Lambda \propto \log \lambda$.
We will show that this extra term will be canceled by the initial field.

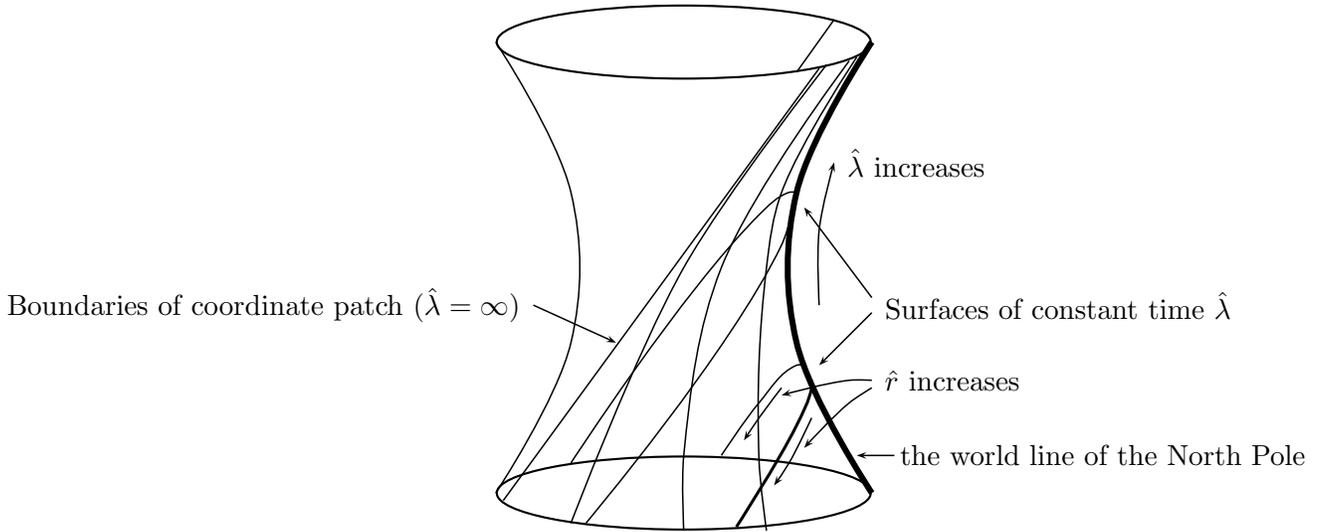
\begin{figure}[t]
\begin{center}
\begin{pspicture}(-4,-8)(4,0)
\pscurve[linewidth=0.2mm]{-}(-2.5,-1)(-1.5,-3)(-1.5,-5)(-2.5,-7)
\pscurve[linewidth=0.8mm]{-}(2.5,-1)(1.5,-3)(1.5,-5)(2.5,-7)
\psline[linewidth=0.2mm](-2.4,-7.1)(1.8,-1.35)
\psline[linewidth=0.2mm](2,-0.7)(1.5,-1.4)
\pscurve[linewidth=0.2mm]{-}(-1.5,-6.6)(1,-3.3)(1.4,-3)(1.5,-3)
\pscurve[linewidth=0.2mm]{-}(1.5,-3)(1.3,-3.7)(-1.3,-7.4)
\pscurve[linewidth=0.2mm]{-}(0.5,-6.5)(1,-5.8)(1.5,-5.3)(1.6,-5.3)
\pscurve[linewidth=0.4mm]{-}(0.7,-7.45)(1.5,-6.1)(1.7,-5.6)(1.7,-5.5)
\pscurve[linewidth=0.2mm]{-}(1.1,-7.5)(1,-6.55)(1.1,-4)(1.3,-3)(2.3,-1.2)
\pscurve[linewidth=0.2mm]{-}(0,-7.5)(0,-6.5)(0.5,-4)(2.2,-1.25)
\pscurve[linewidth=0.2mm]{-}(-1.5,-7.4)(0,-4)(1.9,-1.3)
 \psellipse(0,-1)(2.5,0.5)
 \psellipse(0,-7)(2.5,0.5)
\psline[linewidth=0.2mm]{<-}(1.6,-3.2)(2.5,-4.4)
\psline[linewidth=0.2mm]{<-}(1.8,-5.3)(2.5,-4.6)
    \uput[r](2.5,-4.5){Surfaces of constant time $\hat{\lambda}$}
\pscurve[linewidth=0.2mm]{->}(1.7,-6)(1.4,-6.6)(1.2,-6.9)
\pscurve[linewidth=0.2mm]{->}(1.3,-5.6)(1,-6)(0.8,-6.3)
\pscurve[linewidth=0.2mm]{<-}(1.3,-5.7)(2.2,-5.5)(2.5,-5.5)
\pscurve[linewidth=0.2mm]{<-}(1.6,-6.4)(2.2,-5.8)(2.5,-5.6)
    \uput[r](2.5,-5.5){$\hat{r}$ increases}
    \uput[r](2.7,-6.5){the world line of the North Pole}
\psline[linewidth=0.2mm]{<-}(2.3,-6.5)(2.8,-6.5)
\pscurve[linewidth=0.2mm]{->}(1.8,-4.5)(1.8,-3.5)(2,-2.6)
    \uput[r](2,-2.6){$\hat{\lambda}$ increases}
\psline[linewidth=0.2mm]{<-}(-0.9,-5)(-2,-4.5)
    \uput[l](-2,-4.5){Boundaries of coordinate patch $(\hat{\lambda}=\infty)$}
\end{pspicture}
\end{center}
\caption{Coordinates $(\hat{\lambda},\hat{r},\hat{\theta},\hat{\phi})$ covering the lower half of the manifold.} \label{deSitter:flat2}
\end{figure}

Next we turn our attention to the initial field.  The coordinate system we have used, which covers the causal future of the charge $q$ at the North Pole, is not appropriate because past infinity, where we need to specify the initial data, is not covered at all.  For this reason, we will work in the coordinate system which covers the causal {\em past} of the charge.
Thus, we consider the coordinate system covering the region illustrated in Fig.~\ref{deSitter:flat2}, for which the metric takes the same form as before:
\beq
    ds^2 = \frac{1}{H^2\hat{\lambda}^2}\left(-d\hat{\lambda}^2+d\hat{r}^2+\hat{r}^2d\hat{\Omega}^2\right) \,.
\eeq
The coordinates $\hat{\lambda}$ and $\hat{r}$ are related to $\lambda$ and $r$ for the upper-half spatially-flat coordinate system by
\beqa
    \lambda &=& \frac{\hat{\lambda}}{H^2(\hat{\lambda}^2-\hat{r}^2)} \,, \label{AHtrans1}\\
    r &=& \frac{\hat{r}}{H^2(\hat{\lambda}^2-\hat{r}^2)} \label{AHtrans2}
\eeqa
in the overlapping region.  The angular variables are related trivially as $\theta = \hat{\theta}$ and $\phi = \hat{\phi}$.
The time coordinate $\hat\lambda \in (0,\infty)$ runs forward from the infinite past $(\hat\lambda = 0)$
to the infinite future $(\hat\lambda = \infty)$.
The initial surface will be a hypersurface of constant $\hat\lambda'$, which will be left arbitrary though we are mainly interested in the limit $\hat\lambda'\to 0$.

We let the unprimed coordinates correspond to the point where the field is calculated and the primed coordinates correspond to the point on the initial surface $\Sigma$.
In this coordinate system we have $\hat{\lambda} > \hat{\lambda}'$.
We find from Eq.~(\ref{A_T}) that the input field on the initial surface is given by
\begin{eqnarray}
    A_{\hat\lambda'} & = & -\frac{q}{4\pi}\left( \frac{1}{\hat{r}'} - \frac{1}{\hat{r}'+\hat\lambda'}\right)\,, \label{At} \\
    A_{\hat r'} & = & \frac{q}{4\pi}\frac{1}{\hat{r}'+\hat\lambda'}\,, \label{Ar}
\end{eqnarray}
with $A_{\hat\theta'} = A_{\hat\phi'} = 0$.
The surface element is
\beq
    d\Sigma N_{c'} = \frac{\hat{r}'^2\sin\hat\theta'}{H^4\hat\lambda'^4}d\hat{r}'d\hat\theta'd\hat\phi'\,\delta^{\hat\lambda'}_{c'}\,.
\eeq
We only need the $\hat\lambda'$-component of the conjugate momentum current, $\pi^{\hat\lambda'a'}$. In the Feynman gauge $\zeta=1$ the only nonzero component of the conjugate momentum current is
\beq
    \pi^{\hat\lambda'\hat{r}'} = -g^{\hat\lambda'\hat\lambda'}g^{\hat{r}'\hat{r}'}(\partial_{\hat\lambda'}A_{\hat{r}'} - \partial_{\hat{r}'}A_{\hat\lambda'})
    = -\frac{q}{4\pi} \frac{H^4\hat\lambda'^4}{\hat{r}'^2} \label{pi}.
\eeq
The differential operator $L_{\pi}$ acts on the Green's function as
\begin{eqnarray}
    {(L_{\pi}G)_a}^{c'a'}
    & = & \frac{1}{16\pi}
    \left\{ H^2 \left[ \delta'(z-1) + 2\delta(z-1) \right] g^{c'a'}\partial_a\cos H\mu \right.\nonumber \\
    && \qquad\left. + 2\delta'(z-1) \partial_a \partial^{\,\left[c'\right.}\cos H\mu \cdot \partial^{\left.\,a'\right]}\cos H\mu \right\}\,.
    \label{LG_EM}
\end{eqnarray}
The function $\cos H\mu$ is given by Eq.~(\ref{cosmu}) with all variables replaced by the hatted equivalent. (For example, $\lambda$ is replaced by $\hat\lambda$.)  The following formulas from Ref.~\cite{AllenJacobson} are useful in deriving Eq.~(\ref{LG_EM}):
\begin{eqnarray}
\nabla_{a'}z & = & - H\sqrt{z(1-z)}\,n_{a'}\,, \label{AHAJ1}\\
\nabla_{a'}n_{b'} & = & \frac{H(2z-1)}{2\sqrt{z(1-z)}}(g_{a'b'}-n_{a'}n_{b'})\,,\\
\nabla_{a'}n_a & = & - \frac{H}{2\sqrt{z(1-z)}}(g_{aa'} + n_{a}n_{a'})\,,\\
\nabla_{a'}g_{ab'} & = & H\sqrt{\frac{1-z}{z}}(g_{a'b'}n_a + g_{a'a}n_{b'})\,. \label{AHAJ4}
\end{eqnarray}

\begin{figure}[t]
\begin{center}
\begin{pspicture}(-4,1)(7,9)
\psline[linewidth=1mm](4,3)(4,8)
    \psline[linewidth=0.2mm]{<-}(4,6.5)(4.5,6.5)
    \uput[r](4.5,6.8){world line of a}
    \uput[r](4.5,6.4){static charge}
    \uput[r](4,8.5){spatial origin}
    \uput[r](4,8.1){$\mathbf{x'}=0$}
\psline[linewidth=0.2mm]{->}(3.8,7.5)(3.8,8.5)
    \uput[l](3.8,8){$\hat\lambda$}
\psline[linewidth=0.2mm](4,3)(1,6)
\psline[linewidth=0.2mm, linestyle=dashed](1,6)(0,7)
\psline[linewidth=0.2mm](4,3)(5,4)
\psline[linewidth=0.2mm, linestyle=dashed](5,4)(6,5)
\psellipse[linewidth=0.2mm](-1,3)(3,1)
\psline[linewidth=0.2mm](-3.85,3.3)(-1,6)
\psline[linewidth=0.2mm](1.85,3.3)(-1,6)
\psdots(-1,6)
\psline[linewidth=0.2mm, linestyle=dashed](-1,3)(4,3)
\psline[linewidth=0.2mm, linestyle=dashed](0.7,2.2)(4,3)
\psline[linewidth=0.2mm, linestyle=dashed](-1,3)(0.7,2.2)
\psline[linewidth=0.2mm](0.4,2.35)(0.8,2.45)
\psline[linewidth=0.2mm](0.8,2.45)(1.1,2.3)
\pscurve[linewidth=0.2mm, arrowsize=3pt]{->}(2.9,3.3)(3.05,3.1)(3.2,3)
\pscurve[linewidth=0.2mm, arrowsize=3pt]{<-}(3.4,2.8)(3.7,2.65)(3.9,2.6)
    \uput[r](3.9,2.6){$\hat\Theta$}
\psline[linewidth=0.2mm, linestyle=dashed, arrowsize=4pt]{->}(4,6)(-1,6)
    \uput[u](2,6){$\hat r$}
\psline[linewidth=0.2mm, linestyle=dashed, arrowsize=3pt]{<->}(4.3,3)(4.3,6)
    \uput[r](4.3,5.5){$\hat\lambda-\hat\lambda'$}
\psline[linewidth=0.2mm, linestyle=dashed](-3,2.25)(4,3)
    \psline[linewidth=0.2mm]{<-}(-3,2.25)(-3.3,1.7)
        \rput(-3.4,1.4){$\hat r'^+$}
    \psline[linewidth=0.2mm]{<-}(1.9,2.8)(2.3,2.3)
        \rput(2.6,2.2){$\hat r'^-$}
\psline[linewidth=0.2mm, linestyle=dashed](-3,2.25)(-1,3)
\psline[linewidth=0.2mm, linestyle=dashed](-0.55,2.55)(-1,3)
    \psline[linewidth=0.2mm](-0.8,2.5)(-0.9,2.6)
    \psline[linewidth=0.2mm](-0.4,2.65)(-0.9,2.6)
    \psline[linewidth=0.2mm](-0.4,2.65)(-0.25,2.55)
\psline[linewidth=0.2mm](2.5,3)(2.7,2.85)
    \pscurve[linewidth=0.2mm]{<-}(2.5,2.95)(2.3,2.95)(2.2,3.3)
        \uput[u](2.3,3.3){$\hat\theta'$}
\pscurve[linewidth=0.2mm]{<-}(0.2,2.4)(0.2,2)(0.7,1.7)
    \uput[r](0.7,1.7){$\hat\lambda-\hat\lambda'$}
\psline[linewidth=0.2mm]{->}(-2,3.3)(-1.8,2.7)
    \rput(-2,3.5){$\hat\lambda-\hat\lambda'$}
\psline[linewidth=0.2mm]{<->}(-0.45,2.4)(-2.95,2.15)
\psline[linewidth=0.2mm]{<-}(-1.6,2.3)(-1,1.4)
    \uput[d](-1,1.4){$\sqrt{(\hat\lambda-\hat\lambda')^2-\hat r^2\sin^2\hat\theta'}$}
\psline[linewidth=0.2mm]{<-}(-1.1,6.1)(-1.5,6.5)
    \uput[u](-2.5,6.3){$(\hat\lambda,\hat r,\hat\theta=0)$}
    \uput[u](-2.5,6.9){Field point}
\end{pspicture}
\end{center}
\caption{The causal past of the point $(\hat\lambda,\hat{r},\hat\theta=0)$ in the case
$\hat r>\hat\lambda-\hat\lambda'$.} \label{large r}
\begin{center}
\begin{pspicture}(-5,0)(5,8)
\psline[linewidth=1mm](1.5,2)(1.5,7)
    \psline[linewidth=0.2mm]{<-}(1.5,5.5)(2,5.5)
    \uput[r](2,5.8){world line of a}
    \uput[r](2,5.4){static charge}
    \uput[r](1.5,7.5){spatial origin}
    \uput[r](1.5,7.1){$\mathbf{x'}=0$}
\psline[linewidth=0.2mm]{->}(1.3,6.5)(1.3,7.5)
    \uput[l](1.3,7){$\hat\lambda$}
\psellipse[linewidth=0.2mm](0,2)(3,1)
\psline[linewidth=0.2mm](-2.75,2.4)(0,5)
\psline[linewidth=0.2mm](2.75,2.4)(0,5)
    \psdots(0,5)
\psline[linewidth=0.2mm, linestyle=dashed, arrowsize=4pt]{->}(1.5,5)(0,5)
    \uput[u](0.5,5){$\hat r$}
\psline[linewidth=0.2mm]{<->}(2,2)(2,5)
    \uput[r](2,4){$\hat\lambda-\hat\lambda'$}
\psline[linewidth=0.2mm, linestyle=dashed](1.5,2)(-3,2)
\psline[linewidth=0.2mm, linestyle=dashed](0,2)(-1,1.1)
\psline[linewidth=0.2mm, linestyle=dashed](0,2)(0.4,1.6)
\psline[linewidth=0.2mm]{<->}(0.45,1.5)(-0.9,1)
\psline[linewidth=0.2mm]{<-}(0,1.3)(-0.5,0.3)
    \uput[d](-1.8,0.3){$\sqrt{(\hat\lambda-\hat\lambda')^2-\hat r^2\sin^2\hat\theta'}$}
\psline[linewidth=0.2mm, linestyle=dashed](1.5,2)(-1,1.1)
    \rput(-1.2,0.8){$\hat r'^+$}
\psline[linewidth=0.2mm](0.2,1.55)(0.1,1.65)
\psline[linewidth=0.2mm](0.1,1.65)(0.45,1.8)
\psline[linewidth=0.2mm](0.6,1.65)(0.45,1.8)
\psline[linewidth=0.2mm](1.5,2)(-1,4.5)
\psline[linewidth=0.2mm, linestyle=dashed](-1,4.5)(-1.5,5)
\psline[linewidth=0.2mm](1.5,2)(3.5,4)
\psline[linewidth=0.2mm, linestyle=dashed](3.5,4)(4.5,5)
\pscurve[linewidth=0.2mm](0.7,2)(0.8,1.9)(1,1.8)
\pscurve[linewidth=0.2mm]{<-}(0.8,1.9)(0.5,1.3)(0.7,0.7)
    \uput[d](0.9,0.8){$\hat\theta'$}
\psline[linewidth=0.2mm]{<-}(-0.1,5.1)(-1,6)
    \rput(-2,6.3){$(\hat\lambda,\hat r,\hat\theta=0)$}
    \rput(-2,6.9){Field point}
\pscurve[linewidth=0.2mm]{<-}(-0.6,1.5)(-1.5,1.5)(-2.5,1)
    \rput(-3,0.7){$\hat\lambda-\hat\lambda'$}
\end{pspicture}
\end{center}
\caption{The causal past of the point $(\hat\lambda,\hat r,\hat\theta=0)$ in the case
$\hat r<\hat\lambda-\hat\lambda'$.} \label{small r}
\end{figure}

We specialize to the point $(\hat\lambda,\hat r,\hat{\theta}=0)$ --- $\hat\phi$ is undetermined --- for calculating the field without loss of generality because of the spherical symmetry.
The integral on the initial hypersurface with the time parameter $\hat\lambda'$ in Eq.~(\ref{key formula_original}), which gives the initial field, is over $\hat{r}'$ and $\hat{\theta}'$. (The integration over $\hat{\phi}'$ gives a factor of $2\pi$.)  Due to the causal nature of the retarded Green's function the nonzero contribution comes only from the intersection of $\Sigma$ with the causal past of the point $(\hat\lambda,\hat r,\hat{\theta}=0)$, which is a $3$-dimensional ball, $B$.
There are two cases to be considered separately: i) $\hat r>\hat\lambda-\hat\lambda'$ [$(\hat{\lambda}',\mathbf{0})\notin B$] and ii) $\hat r<\hat\lambda-\hat\lambda'$ [$(\hat{\lambda}',\mathbf{0})\in B$].
For the case $\hat r>\hat\lambda-\hat\lambda'$ (see Fig.~\ref{large r}), for given $\hat{\theta}'$ the range of integration for the radius $\hat r'$ is from $\hat r'^-$ to $\hat r'^+$,
where $\hat r'^{\pm}=\hat r\cos\hat\theta' \pm \sqrt{(\hat\lambda-\hat\lambda')^2-\hat r^2\sin^2\hat\theta'}$.  The variable $\hat{\theta}'$ is integrated from $0$ to $\hat\Theta=\arcsin\left(\frac{\hat\lambda-\hat\lambda'}{\hat r}\right)$.
For the case $\hat r<\hat\lambda-\hat\lambda'$ (see Fig.~\ref{small r}), the integral over $\hat r'$ is from $0$ to $\hat r'^+$ for fixed $\hat{\theta}'$.  Then the angle $\hat\theta'$ is integrated from $0$ to $\pi$.
Using Eqs.~(\ref{G_delta}), (\ref{G_theta}), (\ref{At}), (\ref{Ar}), (\ref{pi}) and (\ref{LG_EM}) in
Eq.~(\ref{key formula_original}),
we obtain the initial field in the part of the spacetime covered by the lower-half spatially-flat coordinate system as
\begin{eqnarray}
    A^{(I)}_{\hat\lambda} &=& -\frac{q}{4\pi}\left(\frac{1}{\hat r}-\frac{1}{\hat r+\hat\lambda}\right)
    \theta(\hat r-\hat\lambda+\hat\lambda')\nonumber \\
     &&        - \frac{q}{12\pi}\left(\frac{1}{\hat\lambda}-\frac{1}{\hat\lambda+\hat r}-\frac{1}{\hat\lambda-\hat r}\right)
        \theta(\hat\lambda-\hat\lambda'-\hat r) \,, \label{AHAIlambda}\\
    A^{(I)}_{\hat r} &=& \frac{q}{4\pi(\hat r+\hat\lambda)}\theta(\hat r-\hat\lambda+\hat\lambda')
    + \frac{q}{12\pi}\left(\frac{1}{\hat\lambda+\hat r}-\frac{1}{\hat\lambda-\hat r}\right)
    \theta(\hat\lambda-\hat\lambda'-\hat r) \,,\label{AHAIr}
\end{eqnarray}
with $A_{\hat\theta}^{(I)} = A_{\hat\phi}^{(I)} = 0$.
The initial field in the region with $\hat{r}>\hat\lambda - \hat\lambda'$ is equal to the field given by Eqs.~(\ref{At}) and (\ref{Ar}) as expected because this region is not influenced by either charge.
In the limit $\hat\lambda'\to 0$ the terms proportional to $\theta(\hat\lambda-\hat\lambda'-r)$ gives the field $A_a^{(I)}$ in the region covered by both $(\hat\lambda,\hat r)$ and $(\lambda,r)$ [see Eqs.~(\ref{AHtrans1}) and (\ref{AHtrans2})].  This field can be expressed in the upper-half coordinates $(\lambda,r)$ as
%
\begin{equation}
    \left. A^{(I)}_{\lambda}\right|_{\rm overlap} =
        - \frac{q}{12\pi\lambda} \,,
\end{equation}
with all other components vanishing.  This term cancels the extra term in Eq.~(\ref{Source:At}), making the sum $A^{(S)}_a+A^{(I)}_a$ agree with the expected field in this region as well.

The calculation in this section shows that the retarded Green's function for the massless vector field obtained from Ref.~\cite{AllenJacobson} correctly reproduces the field compatible with a charge $q$ at the North Pole and a charge $-q$ at the South Pole in the causal past of the charge $q$ at the North Pole.  It is clear that this is also the case in the causal past of the charge $-q$ at the South Pole.  Then, the uniqueness of the solution to the gauge-fixed field equations with given initial data on a $\tau$-constant surface with $\tau < 0$ in the global coordinate system, which is in the causal past of the two charges, implies that the retarded Green's function reproduces the field correctly over the whole spacetime through Eq.~(\ref{key formula}).

\section{Gravitational field in de~Sitter spacetime} \label{case3}

In this section we reproduce the linearized gravitational field compatible with two point masses $M$, one at the North Pole and the other at the South Pole of de~Sitter spacetime, using the retarded Green's function obtained from the two-point function found in Ref.~\cite{Higuchi2}.  (One cannot have a single mass point in de~Sitter spacetime because the total conserved charge corresponding to a de~Sitter boost symmetry must vanish.  This fact is related to the linearization instabilities of the Einstein equations about any spatially compact spacetime with Killing vectors such as de~Sitter spacetime~\cite{BrillDeser,FischerMarsden}.)

\subsection{Linearized gravity for a point mass in a covariant gauge} \label{AHgravfield}

We first construct the field to be reproduced starting from the Schwarzschild-de~Sitter solution in the static coordinate system:
\beq
    ds^2 = -\left(1-H^2R^2- \frac{2M}{R}\right)dT^2 + \left(1-H^2R^2-\frac{2M}{R}\right)^{-1}dR^2
           + R^2d\Omega^2\,. \label{Schwarzschild deSitter}
\eeq
As is well known, this solution represents two black holes of equal masses $M$ in de~Sitter spacetime.  If we write this metric as $\tilde{g}_{\mu\nu} = g_{ab} + h_{ab}$, where $g_{ab}$ is the background de~Sitter metric (\ref{AHstatmet}), then the linear term $h_{ab}$ is singular at the horizon $R=H^{-1}$.  However, after the coordinate transformation given by
\begin{eqnarray}
    R &\to& \left(1-\frac{2}{3}MH\right)R - \frac{M}{3} \,, \label{new:r} \\
    T &\to& (1+2MH)T \,, \label{new:t}
\end{eqnarray}
we find
\begin{eqnarray}
    h_{TT} &=& \frac{2M}{R}(1-HR)\left(1-HR-\frac{4}{3}H^2R^2\right) \,, \label{AHh00} \\
    h_{RR}
        &=& \frac{2M}{3R(1-H^2R^2)}\left(1+\frac{2}{1+HR}\right) \,, \label{AHh11}\\
    h_{ij} &=& -\frac{2}{3}MR(1+2HR)\eta_{ij} \,, \label{AHh33}
\end{eqnarray}
where $i,j = \theta, \phi$ and where $\eta_{ij}$ is the standard metric on the unit $2$-sphere.
The singularity of the component $h_{RR}$ at the horizon $R=H^{-1}$ is a coordinate singularity, and this field is non-singular at the horizon as one can see by expressing $h_{ab}$ in the global coordinate system.  The expression of $h_{ab}$ in global coordinates also makes it clear that there are two mass points, one at the North Pole and the other at the South Pole.

The trace and divergence are
\beqa
    {h^a}_a & = & -\frac{4M}{3R} \,,\\
    {\nabla_{b}h^{b}}_{a} & = & \frac{10M}{3R^2}\delta^R_{a} \,.
\eeqa
Hence $h_{ab}$ satisfies the gauge condition
\beq
\nabla_b{h^b}_a - \frac{1+\beta}{\beta}\nabla_a h   = 0 \label{AHgaugecond}
\eeq
with $\beta = \frac{2}{3}$. This gauge condition is exactly the one for which an explicit form of the Feynman propagator is given in Ref.~\cite{Higuchi2}.

It is possible to make the field $h_{ab}$ traceless and divergence-free at $R\neq 0$ by a further gauge transformation,
\beq
R \to R - \frac{M}{3H^2R^2}(1-H^2 R^2).
\eeq
However, the resulting $RR$-component will behave like $R^{-3}$ at $R=0$.  We do not adopt this solution for this reason and will work with that found above.

In the conformally-flat coordinates covering the upper half of the spacetime we find
\beqa
h_{\lambda\lambda} & = & \frac{4M}{3H}\left[\frac{2}{r(r+\lambda)} - \frac{1}{(r+\lambda)^2} - \frac{1}{2r\lambda}\right]\,,\label{h00} \\
h_{rr} & = & \frac{4M}{3H}\left[\frac{2}{r(r+\lambda)} - \frac{1}{(r+\lambda)^2} + \frac{2}{\lambda^2}-\frac{1}{2r\lambda}\right]\,,\label{h11}\\
h_{\lambda r} & = & \frac{4M}{3H}\left[\frac{2}{r(r+\lambda)} - \frac{1}{(r+\lambda)^2} - \frac{2}{r\lambda}\right]\,,\label{h22}\\
h_{ij} & = & - \frac{4M}{3H}\left(\frac{r}{2\lambda} +
\frac{r^2}{\lambda^2}\right)\eta_{ij}\,,\,\,i,j=\theta,\phi\,,\label{h33}
\eeqa
with all other components vanishing. We reproduce this field using the retarded Green's function obtained from the result of Ref.~\cite{Higuchi2} with $\beta = \frac{2}{3}$.


\subsection{The retarded Green's function for linearized gravity}

The Lagrangian density of pure gravity with positive cosmological constant $3H^2$ is given by
\beq
    \mathcal{L}_{\text{full}} = (16\pi)^{-1}\sqrt{-\tilde{g}}(R-6H^2) \,, \label{Lagrangian full}
\eeq
where $\tilde{g}$ is the full metric and $R$ is the Ricci scalar.
We write the metric as $\tilde{g}_{ab} = g_{ab} + h_{ab}$, where $g_{ab}$ is the background de~Sitter metric,
expand the Lagrangian (\ref{Lagrangian full}) to second order in $h_{ab}$, and then include the gauge fixing term,
\beq
    \mathcal{L}_{\text{gf}} = -\frac{\sqrt{-g}}{32\pi \alpha} \( \nabla_ah^{ab} - \frac{1+\beta}{\beta}\nabla^bh \)
                        \( \nabla^ch_{cb} - \frac{1+\beta}{\beta}\nabla_bh \)\,,
\eeq
where $\alpha$ and $\beta$ are the gauge parameters.  Specializing to the case with $\beta = \frac{2}{3}$, we find
the Euler-Lagrange field equations to be
\beqa
    -{L_{ab}}^{cd}h_{cd} &=& \tfrac{1}{2}\square h_{ab} - \(\frac{1}{2}-\frac{1}{2\alpha}\)(\nabla_a\nabla_c{h^c}_b
                            + \nabla_b\nabla_c{h^c}_a)
                    + \[ \frac{1}{2}-\frac{5}{2\alpha} \]\nabla_a\nabla_bh \nol \\
                    && + \[\frac{25}{4\alpha} - \frac{1}{2} \]g_{ab}\square h
                    + \tfrac{1}{2}g_{ab} \( 1-\frac{5}{\alpha} \)\nabla_c\nabla_dh^{cd}
                    - H^2(h_{ab}+\tfrac{1}{2}g_{ab}h)\nonumber \\
                    &  = &  0\,. \label{Lh}
\eeqa
(The definition of ${L_{ab}}^{cd}$ differs from that in Ref.~\cite{Higuchi2} by a minus sign.)  The Green's function of the operator ${L_{ab}}^{cd}$ given by Eq.~(\ref{Lh}) was calculated in Ref.~\cite{Higuchi2} by analytically continuing the corresponding Green's function on $S^4$ of radius $H^{-1}$.  The corresponding Feynman propagator $Q_{aba'b'}(x,x')$ satisfies
\beq
    L_x^{abcd}Q_{cda'b'}(x,x') = -i{\delta^{ab}}_{a'b'}(x,x') \,, \label{-LG}
\eeq
where the subscript $x$ in $L_x^{abcd}$ indicates that the differential operator $L^{abcd}$ acts at $x$ rather than $x'$.
Here the $\delta$-function is defined by
\beq
    \int \sqrt{-g(x')}\,d^4x'\,\delta_{aba'b'}(x,x')f^{a'b'}(x') = f_{ab}(x)
\eeq
for any smooth compactly-supported symmetric tensor $f_{ab}$.
As in the original graviton two-point function in the gauge $\alpha=1$ and $\beta = -6$~\cite{AllenTuryn}, the covariant graviton propagator in the gauge $\beta=\frac{2}{3}$
can be expressed in terms of bi-tensors formed using the five basic (bi-)tensors $g_{ab},g_{a'b'},g_{aa'},n_a,n_{a'}$.
They are
\begin{eqnarray}
    O^{(1)}_{aba'b'} &=& g_{ab}g_{a'b'} \,, \\
    O^{(2)}_{aba'b'} &=& g_{aa'}g_{bb'} + g_{ab'}g_{ba'} \,, \\
    O^{(3)}_{aba'b'} &=& g_{ab}n_{a'}n_{b'} + g_{a'b'}n_an_b \,, \\
    O^{(4)}_{aba'b'} &=& g_{aa'}n_bn_{b'} + g_{ab'}n_bn_{a'} + g_{ba'}n_an_{b'} + g_{bb'}n_an_{a'} \,, \\
    O^{(5)}_{aba'b'} &=& n_an_bn_{a'}n_{b'} \,,
\end{eqnarray}
where $g_{ab'}$ and $n_a$ are defined by Eqs.~(\ref{n}) and (\ref{prllprop}), and where $g_{ab}$ and $g_{a'b'}$ are the metric tensors at $x$ and $x'$, respectively.
The covariant graviton propagator is
\beq
    Q_{aba'b'}(x,x') = \frac{H^2}{16\pi^2} \sum_{k=1}^5 f_{(k)}(z) \,\, O_{aba'b'}^{(k)}\,, \label{graviton_propagator}
\eeq
where $f_{(k)}(z)$ are scalar functions of $z(x,x')$ given by
\begin{eqnarray}
    f_{(1)}(z) &=&  \frac{22\alpha}{15}-\frac{4}{9}
                    + \left(\frac{8}{3}-\frac{17\alpha}{45}\right)\frac{1}{z}
                    + \left(\frac{\alpha}{45}-\frac{4}{9}\right)\frac{2}{z^2}
                    + \left(\frac{1}{3}-\frac{4\alpha}{9}\right)\frac{1}{1-z} \nol \\
                    && + \left[\frac{4\alpha}{5}-\frac{4}{3}
                    + \left(\frac{\alpha}{9}-1\right)\frac{2}{z}
                    + \left(\frac{14}{9}-\frac{\alpha}{5}\right)\frac{2}{z^2}
                    + \left(\frac{\alpha}{45}-\frac{4}{9}\right)\frac{2}{z^3}\right]\log(1-z) \,,\quad\\
    f_{(2)}(z) &=&  \frac{2}{3}-\frac{13\alpha}{5}
                    - \left(\frac{2\alpha}{45}+\frac{1}{3}\right)\frac{1}{z}
                    + \left(\frac{\alpha}{45}-\frac{4}{9}\right)\frac{2}{z^2}
                    + \left(\frac{5\alpha}{9}-\frac{4}{3}\right)\frac{1}{1-z} \nol \\
                    && + \left[2-\frac{6\alpha}{5}
                    + \left(\frac{1}{9}-\frac{\alpha}{15}\right)\frac{1}{z^2}
                    + \left(\frac{\alpha}{45}-\frac{4}{9}\right)\frac{2}{z^3}\right]\log(1-z) \,,\\
    f_{(3)}(z) &=&  \frac{4}{9}-\frac{2\alpha}{3}
                    + \left(\frac{31\alpha}{45}-4\right)\frac{2}{z}
                    + \left(\frac{4}{3}-\frac{\alpha}{15}\right)\frac{4}{z^2}
                    + \frac{8\alpha}{9}\frac{1}{1-z} \nol \\
                    && + (1-z)\left[ \left(\frac{\alpha}{5}-\frac{1}{3}\right)\frac{4}{z}
                    + \left(\frac{7\alpha}{45}-\frac{2}{3}\right)\frac{8}{z^2}
                    + \left(\frac{4}{3}-\frac{\alpha}{15}\right)\frac{4}{z^3}\right]\log(1-z) \,,\\
    f_{(4)}(z) &=&  \frac{2}{3}-\frac{16\alpha}{5} + \frac{6\alpha}{5}z
                    + \left(\frac{7\alpha}{45}-\frac{2}{3}\right)\frac{4}{z}
                    + \left(\frac{4}{9}-\frac{\alpha}{45}\right)\frac{8}{z^2}
                    + \left(\frac{\alpha-33}{9}\right)\frac{1}{1-z} \nol \\
                    && + (1-z)\left[ \left(\frac{3\alpha}{5}-1\right)\frac{2}{z}
                    + \left(\frac{\alpha}{15}-\frac{1}{9}\right)\frac{8}{z^2}
                    + \left(\frac{4}{9}-\frac{\alpha}{45}\right)\frac{8}{z^3}\right]\log(1-z) \,,\\
    f_{(5)}(z) &=&  \frac{8}{9}-\frac{152\alpha}{15}+\frac{24\alpha}{5}z
                    + \left(\frac{5}{3}-\frac{4\alpha}{15}\right)\frac{8}{z}
                    + \left(\frac{\alpha}{45}-\frac{4}{9}\right)\frac{16}{z^2}
                    - \left(\frac{1}{3}+\frac{\alpha}{9}\right)\frac{32}{1-z} \nol \\
                    && + (1-z)^2\left[ \left(\frac{1}{3}-\frac{\alpha}{5}\right)\frac{8}{z^2}
                    + \left(\frac{\alpha}{45}-\frac{4}{9}\right)\frac{16}{z^3}\right]\log(1-z) \,.
\end{eqnarray}
In finding the retarded Green's function it is important to note that the products $n_an_{b}$ and $n_{a}n_{a'}$ contribute a factor $(1-z)^{-1}$ [see Eq.~(\ref{n})], which is singular as $z\to 1$. (We did not face this issue in the electromagnetic case because the bi-vector $n_an_{a'}$ was multiplied by $1-z$ in the two-point function.)  We note that there is no term of the singularity structure of the form $(1-z)^{-n}\log(1-z)$, $n \geq 1$, in $Q_{aba'b'}(x,x')$.  Such a term would have caused difficulties in finding the retarded Green's function.


It is convenient to define
\beqa
    {C^{(1)}}_{abc'd'} &=& g_{ab}\,g_{c'd'} \,, \label{AHC1}\\
    {C^{(2)}}_{abc'd'} &=& \frac{1}{H^2} \left( g_{ab}\,\partial_{c'}\cos H\mu\cdot\partial_{d'}\cos H\mu+g_{c'd'}\,\partial_a\cos H\mu\cdot\partial_b\cos H\mu \right) \,, \qquad \label{AHC2}\\
    {C^{(3)}}_{abc'd'} &=& \frac{1}{H^4} \left( \partial_a\cos H\mu\cdot\partial_b\cos H\mu\cdot\partial_{c'}\cos H\mu\cdot\partial_{d'}\cos H\mu \right) \,. \label{AHC3}
\eeqa
The bi-tensors $C^{(n)}_{ac'bd'}$, $n=1,2,3$, are defined by swapping $b$ and $c'$ in these formulas.
The retarded Green's function $G_{aba'b'}(x,x')$ can be obtained in the same manner as in the electromagnetic case by noting
\beq
    G_{aba'b'}(x,x') = -i\theta(\tau-\tau')\left[Q_{aba'b'}(z+i\epsilon) - Q_{aba'b'}(z-i\epsilon)\right] \,,
\eeq
and using the Sokhotski formula (\ref{Sokhotski1}) and its derivatives,
\beq
    \left(\frac{1}{1-z \pm i\epsilon}\right)^{n+1} = \frac{1}{n!}\left[\frac{\partial^n}{\partial z^n}\left({\rm P}\,\frac{1}{1-z}\right) \pm (-1)^{n+1}i\pi \delta^{(n)}(1-z)\right]
    \,, \label{Sokhotski3}
\eeq
and Eq.~(\ref{Sokhotski4}).
We obtain the retarded Green's function in this manner as
\beqa
    G_{aba'b'} &=& \frac{H^2}{8\pi} \left[ G^{(0)}_{aba'b'}\delta(1-z) + G^{(1)}_{aba'b'}\delta'(1-z)
                    + G^{(2)}_{aba'b'}\delta''(1-z)\right.\nonumber \\
                &&  \left. \qquad + G^{(\theta)}_{aba'b'}\theta(z-1) \right]\,,  \label{AHGdelta}
\eeqa
where
\beqa
    G^{(0)}_{aba'b'}
        &=& \left(\frac{1}{3}-\frac{4\alpha}{9}\right) {C^{(1)}}_{aba'b'}
        + \left(\frac{5\alpha}{9}-\frac{4}{3}\right)\left( {C^{(1)}}_{aa'bb'} + {C^{(1)}}_{a'bb'a} \right)
        + \frac{5}{9}\left(\frac{3\alpha}{5}-1\right){C^{(2)}}_{aba'b'} \nol \\
        && - \frac{1}{36}(19+13\alpha)\left({C^{(2)}}_{ab'ba'} + {C^{(2)}}_{aa'bb'}\right)
        - \frac{7}{6}\left(1+\frac{5\alpha}{3}\right)\,{C^{(3)}}_{aba'b'}\,, \\
    G^{(1)}_{aba'b'}
        &=& - \frac{2\alpha}{9}{C^{(2)}}_{aba'b'}
        + \left( \frac{11}{12} - \frac{\alpha}{36} \right) \left( {C^{(2)}}_{ab'ba'}+{C^{(2)}}_{aa'bb'} \right)
        + \frac{8}{9}(1+\alpha) {C^{(3)}}_{aba'b'} \,, \\
    G^{(2)}_{aba'b'}
        &=& - \left(\frac{1}{3}+\frac{\alpha}{9}\right)\,{C^{(3)}}_{aba'b'} \,, \\
    G^{(\theta)}_{aba'b'}
        &=& (2z-1)\left[ \left(1-\frac{3\alpha}{5}\right)\frac{2}{3z}
        + \left(\frac{4}{3}-\frac{14\alpha}{45}\right)\frac{1}{z^2}
        + \left(\frac{2\alpha}{45}-\frac{8}{9}\right)\frac{1}{z^3}\right] {C^{(1)}}_{aba'b'} \nol \\
        && + \left[ \left(\frac{3\alpha}{5}-1\right)\left(2+\frac{1}{9z^2}\right)
        + \left(\frac{8}{9}-\frac{2\alpha}{45}\right)\frac{1}{z^3} \right]
        \left({C^{(1)}}_{aa'bb'} + {C^{(1)}}_{a'bb'a}\right) \nol \\
        && + \left[ \left(1-\frac{3\alpha}{5}\right)\frac{1}{3z^2}
        + \left(\frac{4}{3}-\frac{14\alpha}{45}\right)\frac{1}{z^3}
        + \left(\frac{\alpha}{15}-\frac{4}{3}\right)\frac{1}{z^4} \right] {C^{(2)}}_{aba'b'} \nol \\
        && + \left[ \left(1-\frac{3\alpha}{5}\right)\left(\frac{1}{2z^2}+\frac{2}{9z^3}\right)
        + \left(\frac{2\alpha}{45}-\frac{8}{9}\right)\frac{1}{z^4} \right]
        \left( {C^{(2)}}_{ab'a'b} + {C^{(2)}}_{aa'bb'} \right) \nol \\
        && + \left[ \left(\frac{3\alpha}{5}-1\right)\frac{1}{6z^4}
        + \left(\frac{4}{9}-\frac{\alpha}{45}\right)\frac{1}{z^5} \right]{C^{(3)}}_{aba'b'} \,.
\eeqa


\subsection{The source field}

The formula~(\ref{key formula}) for linearized gravity is
\beq
h_{ab}(x) = h_{ab}^{(S)}(x) + h_{ab}^{(I)}(x)\,, \label{key formula:gravity}
\eeq
where
\beqa
    h_{ab}^{(S)}(x) &=& 8\pi \int_{D^+(\Sigma)} d^4x' \, \sqrt{-g'} \, G_{aba'b'}(x,x')T^{a'b'}(x')\,,
    \label{AHsource:gravity}\\
    h_{ab}^{(I)}(x) & = & \int_{\Sigma}d\Sigma_{c'} \left[ \pi^{c'a'b'}(x') G_{aba'b'}(x,x')
                - h_{a'b'}(x') {\left( L_{\pi}G \right)_{ab}}^{c'a'b'}(x,x') \right] \,.
                \label{key formula:initial}
\eeqa
We explicitly show that this formula reproduces the field $h_{ab}$ in Sec.~\ref{AHgravfield} compatible with two mass points $M$ at the North and South Poles.  In the conformally-flat coordinate system covering the upper half of the spacetime, the corresponding stress-energy tensor is proportional to $\delta^3(\mathbf{x})$, representing the mass point at the North Pole.  It can readily be found by using the conservation law $\nabla_a T^{ab} = 0$ and dimensional analysis as
\beq
    T^{\lambda\lambda} = MH^5\lambda^5\delta^3(\bf{x}) \,, \label{T}
\eeq
with all other components vanishing.

The source field (\ref{AHsource:gravity}) is found using Eqs.~(\ref{AHGdelta}) and (\ref{T}) as
%
\begin{eqnarray}
    h^{(S)}_{ab}(x)
    &=& MH^3 \int_0^{\infty} d\lambda' \, \lambda' \left[ G^{(0)}_{ab\lambda'\lambda'}\delta(1-z)
    + G^{(1)}_{ab\lambda'\lambda'}\delta'(1-z) \right. \nol \\
    && \qquad\qquad + \left. \left. G^{(2)}_{ab\lambda'\lambda'}\delta''(1-z)
    + G^{(\theta)}_{ab\lambda'\lambda'}\theta(z-1) \right] \right|_{r',\theta'=0} \,.
\end{eqnarray}
Again we cut off the stress-energy tensor at $\lambda'=\lambda_0$ by multiplying it by $\theta(\lambda_0-\lambda')$ and let $\lambda_0\to \infty$ in the end. By integrating by parts to reduce the order of derivative of the delta functions and letting $\int_0^{\lambda_0}d\lambda'\theta(z-1) = \int_{\lambda+r}^{\lambda_0}d\lambda'$, we obtain the $\lambda\lambda$-component as
\begin{eqnarray}
    h^{(S)}_{\lambda\lambda}
        &=& MH^3 \left\{ \int_0^{\infty} d\lambda' \, \left[ \lambda' G^{(0)}_{\lambda\lambda\lambda'\lambda'}
        + \frac{\partial}{\partial\lambda'}
        \left( \lambda' G^{(1)}_{\lambda\lambda\lambda'\lambda'} \left(\frac{\partial z}{\partial \lambda'}\right)^{-1} \right)
        \right.\right. \nol \\
        && \qquad \left.\left. + \frac{\partial}{\partial\lambda'} \left( \frac{\partial}{\partial\lambda'}
        \left( \lambda' G^{(2)}_{\lambda\lambda\lambda'\lambda'} \left(\frac{\partial z}{\partial \lambda'}\right)^{-1} \right)
        \left(\frac{\partial z}{\partial\lambda'}\right)^{-1} \right) \right]\delta(1-z) \right. \nol \\
        && \qquad + \left. \int_{\lambda+r}^{\lambda_0} d\lambda' \, \lambda' G^{(\theta)}_{\lambda\lambda\lambda'\lambda'} \right\} \nol \\
        &=& \frac{4M}{3H} \left[ \frac{2}{r(r+\lambda)}
        - \frac{1}{(r+\lambda)^2}
        - \frac{1}{2r\lambda}
    	+ \frac{9\alpha}{5\lambda^2} \right]\theta(\lambda_0-\lambda-r) \,.
\end{eqnarray}
The other components can be found in a similar manner as
\begin{eqnarray}
    h^{(S)}_{rr} &=& \frac{4M}{3H} \left[ \frac{2}{r(r+\lambda)}
    - \frac{1}{(r+\lambda)^2}
    - \frac{1}{2\,r\lambda}
    + \left( 2 + \frac{3\alpha}{5}\right)\frac{1}{\lambda^2} \right]\theta(\lambda_0-\lambda-r) \,, \\
    h^{(S)}_{r\lambda}
    &=& \frac{4M}{3H} \left[ \frac{2}{r(r+\lambda)}
    - \frac{1}{(r+\lambda)^2} - \frac{2}{r\lambda} - \frac{9\alpha r}{10\lambda^3} \right]\theta(\lambda_0-\lambda-r) \,, \\
    h^{(S)}_{ij} &=& -\frac{4M}{3H} \left[\frac{r}{2\lambda}+\left( 1- \frac{3\alpha}{5}\right)\frac{r^2}{\lambda^2}\right]\eta_{ij}\theta(\lambda_0-\lambda-r) \,,
\end{eqnarray}
with the remaining components vanishing.  The source field $h^{(S)}_{ab}$ agrees with the field we found earlier given by Eqs.~(\ref{h00})-(\ref{h33}) if we let $\alpha = 0$.  We also find that the $\alpha$-dependent contribution can be expressed as $\nabla_a \Lambda_b + \nabla_b \Lambda_a$ where
\beqa
\Lambda^\lambda & = & - \frac{6\alpha MH\lambda}{5} \log H\lambda\,,\\
\Lambda^r & = & \frac{2\alpha MH r}{5}(1-3\log H\lambda)\,,
\eeqa
with $\Lambda^\theta = \Lambda^\phi = 0$.  Thus, the retarded Green's function reproduces the field correctly in the causal future of the point mass up to a linear gauge transformation.  The source field with $\alpha\neq 0$ does not satisfy the gauge condition:
\beq
F^a \equiv \nabla_b h^{ba} - \frac{5}{2}\nabla^a h =
\begin{cases} - 6\alpha MH^3\lambda &\,,a=\lambda\,, \\
- 6\alpha MH^3r &\,,a=r\,,\\
\quad 0 &\,,a=\theta,\phi\,.\end{cases}
\eeq
The vector $F^a$ is a Killing vector, and as a result the contribution of the gauge-fixing term to the field equations,
$\alpha^{-1}(\nabla_a F_b + \nabla_b F_a - 5g_{ab}\nabla_c F^c)$, vanishes. Since this could not be the case if the spacetime had no Killing vector, we expect that the source field by itself would not satisfy the linearized Einstein equations for $\alpha\neq 0$ in a spacetime without a Killing vector.

In the next section we will show, with $\alpha=\frac{5}{3}$, that the initial field given by Eq.~(\ref{key formula:initial}) reproduces the field in Eqs.~(\ref{h00})-(\ref{h33}) correctly at points which are not in the causal future of the mass points and exactly cancels out the extra pure-gauge term from the source field in their causal future.


\subsection{The initial field}

The calculation of the initial field is rather complicated, and we specialize to the gauge $\alpha=\frac{5}{3}$, which somewhat simplifies the retarded Green's function.
Again we use the conformally-flat coordinate system $(\hat{\lambda},\hat{r},\hat{\theta},\hat{\phi})$ covering the causal past of the point mass at the North Pole.
We let the input field $h_{a'b'}$ on the initial hypersurface $\hat\lambda = \hat\lambda'$ be equal to  Eq.~(\ref{h00})-(\ref{h33}) with each variable changed to its hatted equivalent.

Let us first find the conjugate momentum current $\pi^{cab}$.  As in the electromagnetic case we denote the coordinates at the initial hypersurface by primed variables, $(\hat\lambda',\hat r', \hat\theta', \hat\phi')$, and those at the point where the field is calculated by unprimed variables, $(\hat\lambda,\hat r, \hat\theta, \hat\phi)$.
The Lagrangian density for linearized gravity with gauge-fixing term with $\alpha = \frac{5}{3}$ and $\beta = \frac{2}{3}$ can be written as
\beq
    \mathcal{L} = \frac{\sqrt{-g}}{2} \left[ T^{abcdef}\nabla_ah_{bc}\nabla_dh_{ef} + S^{abcd}H^2 h_{ab}h_{cd} \right] \,.
\eeq
where
\beqa
T^{abcdef} & = & \frac{2}{5}g^{ab}g^{cf}g^{de}
                - \frac{1}{2} g^{ad}g^{be}g^{cf} - \frac{13}{4} g^{ad}g^{bc}g^{ef}
                + g^{ac}g^{bd}g^{ef} + g^{df}g^{ae}g^{bc}\,,\\
S^{abcd} & = & -  g^{ac}g^{bd} + \frac{1}{2} g^{ab}g^{cd}\,.
\eeqa
The conjugate momentum current is given by
\begin{eqnarray}
    \pi^{cab} &=& \frac{1}{\sqrt{-g}}\frac{\partial\mathcal{L}}{\partial(\nabla_c h_{ab})} \nonumber \\
        &=& \tfrac{1}{2}\(T^{cabdef}+T^{cbadef}\) \nabla_d h_{ef} \nonumber \\
        &=& - \tfrac{1}{2}\nabla^c h^{ab} - \tfrac{3}{4}g^{ab}\nabla^c h
            + 2 g^{c(a}\nabla^{b)} h\,,
\end{eqnarray}
where we have used the covariant gauge condition to replace $\nabla_b {h^b}_a$ by $\frac{5}{2}\nabla_a h$.
As for the conjugate momentum current we only need the components $\pi^{\hat\lambda' a'b'}$.   They are
\begin{eqnarray}
    \pi^{\hat\lambda'\hat\lambda'\hat\lambda'}
        &=& \frac{4MH^5}{3}\hat\lambda'^4 \left[ \frac{\hat\lambda'^2}{(\hat\lambda'+\hat r')^3}
        - \frac{1}{\hat\lambda'+\hat r'} -\frac{1}{2\hat r'} \right]\,, \label{AHpi1}\\
    \pi^{\hat\lambda' \hat r'\hat r'}
        &=& \frac{4MH^5}{3}\hat\lambda'^4 \left[ \frac{\hat\lambda'^2}{(\hat\lambda'+\hat r')^3} -
        \frac{1}{\hat \lambda'+\hat r'} \right] \,, \\
    \pi^{\hat\lambda'\hat\lambda' \hat r'}
        &=& \frac{4MH^5}{3}\hat\lambda'^4\left[ - \frac{\hat\lambda'^2}{(\hat\lambda'+\hat r')^3}
        + \frac{1}{\hat\lambda'+\hat r'} -\frac{\hat\lambda'}{\hat r'^2} \right] \,, \\
    \pi^{\hat\lambda'i'j'}
        &=& -\frac{4MH^5\hat\lambda'^4}{3\hat r'^3}\eta^{i'j'} \,,\,\,i',j'=\hat\theta',\hat\phi'\,, \label{AHpi3}
\end{eqnarray}
Next, we evaluate
\beqa
    {\( L_{\pi}G \)_{ab}}^{c'a'b'} &=& \tfrac{1}{2}\(T^{c'a'b'd'e'f'}+T^{c'b'a'd'e'f'}\)\nabla_{d'}G_{abe'f'} \nol \\
        &=& \tfrac{2}{5} g^{c'(a'}\nabla_{d'}{G_{ab}}^{b')d'}
        - \tfrac{1}{2}\nabla^{c'}{G_{ab}}^{a'b'}
        + g^{c'(a'}\nabla^{b')}G_{ab}\nonumber \\
        &&
        + g^{a'b'}\nabla_{d'}{G_{ab}}^{d'c'}
        - \tfrac{13}{4}g^{a'b'}\nabla^{c'}G_{ab} \,, \label{LG}
\eeqa
where $G_{ab} = {{G_{ab}}^{d'}}_{d'}$.  Recall we have expanded $G_{abc'd'}$ in terms of the tensors $C^{(i)}_{abc'd'}$ and $C^{(i)}_{ac'bd'}$, $i=1,2,3$, defined by Eqs.~(\ref{AHC1})-(\ref{AHC3}).  We need the covariant derivatives of these tensors in order to find ${\( L_{\pi}G \)_{ab}}^{c'a'b'}$.  These are evaluated in Appendix~\ref{AppendixB}.

As in the electromagnetic case we group together the terms proportional to $\delta(1-z)$, $\delta'(1-z)$, $\delta''(1-z)$, $\delta'''(1-z)$ and $\theta(z-1)$ in ${\( L_{\pi}G \)_{ab}}^{c'a'b'}$ as
\beqa
    {\left(L_{\pi}G\right)_{ab}}^{c'a'b'}
       & = & \frac{H^3}{8\pi} \left\{ {S^{(0)}_{ab}}^{c'a'b'}\delta(1-z)+{S^{(1)}_{ab}}^{c'a'b'}\delta'(1-z)
        + {S^{(2)}_{ab}}^{c'a'b'}\delta''(1-z)\right.\nonumber \\
        && \qquad \left. + {S^{(3)}_{ab}}^{c'a'b'}\delta'''(1-z)+ {{S^{(\theta)}}_{ab}}^{c'a'b'}\theta(z-1) \right\} \,, \label{AHG}
\eeqa
where
\beqa
    {{S^{(0)}}_{ab}}^{c'a'b'}
    &=& \frac{4}{9}\sqrt{z(1-z)}\left\{ 11g^{c'a'}{g_{(a}}^{b'}n_{b)}
    - \frac{25}{3}{g_{(a}}^{c'}{g_{b)}}^{a'}n^{b'}
    - \frac{11}{2}g^{a'b'}{g_{(a}}^{c'}n_{b)}
    - g_{ab}g^{c'a'}n^{b'} \right. \nol \\
    && + \frac{11}{6}{g_{(a}}^{a'}{g_{b)}}^{b'}n^{c'}
    + \frac{41}{4}g^{a'b'}g_{ab}n^{c'}
    - z(1-z) \left[ \frac{165}{2}g^{a'b'}n_an_bn^{c'}
    + \frac{112}{3}g^{c'a'}n_{a}n_{b}n^{b'} \right. \nol \\
    && \left. + \frac{130}{3}{g_{(a}}^{c'}n_{b)}n^{a'}n^{b'}
    + \frac{152}{3}{g_{(a}}^{a'}n_{b)}n^{b'}n^{c'}
    + \frac{11}{6}g_{ab}n^{a'}n^{b'}n^{c'} \right] \nol \\
    && \left. - 274z^2(1-z)^2 n_an_bn^{a'}n^{b'}n^{c'} \right\} \,, \label{S1}\\
    {{S^{(1)}}_{ab}}^{c'a'b'}
    &=& \frac{4}{9}\sqrt{z(1-z)}\left\{ \frac{47}{6}\left( g^{c'a'}{g_{(a}}^{b'}n_{b)} + {g_{(a}}^{c'}{g_{b)}}^{a'}n^{b'} \right)
    - \frac{14}{3}\left(g_{ab}g^{c'a'}n^{b'}+g^{a'b'}{g_{(a}}^{c'}n_{b)}\right) \right. \nol \\
    && + \frac{11}{12}{g_{(a}}^{a'}{g_{b)}}^{b'}n^{c'}
    - \frac{35}{12}g^{a'b'}g_{ab}n^{c'}
    + z(1-z) \left[ \frac{247}{6}g^{a'b'}n_an_bn^{c'}
    + \frac{41}{3}g^{c'a'}n_an_bn^{b'} \right. \nol \\
    && \left.\left. + 27{g_{(a}}^{c'}n_{b)}n^{a'}n^{b'}
    + 36{g_{(a}}^{a'}n_{b)}n^{b'}n^{c'}
    + 202z(1-z)n_an_bn^{a'}n^{b'}n^{c'} \right]\right\} \,, \label{S2}\\
    {{S^{(2)}}_{ab}}^{c'a'b'}
    &=& \frac{4}{27}\sqrt{z(1-z)}z(1-z)\left\{ -\frac{17}{2}g^{a'b'}n_an_bn^{c'}
    - 28\left(g^{c'a'}n_an_bn^{b'} + {g_{(a}}^{c'}n_{b)}n^{a'}n^{b'}\right) \right. \nol \\
    && \left. - 47{g_{(a}}^{a'}n_{b)}n^{b'}n^{c'}
    + 5g_{ab}n^{a'}n^{b'}n^{c'}
    - 184z(1-z) n_an_bn^{a'}n^{b'}n^{c'} \right\} \,, \label{S3}\\
    {{S^{(3)}}_{ab}}^{c'a'b'}
    &=& \frac{112}{27}z^2(1-z)^2\sqrt{z(1-z)} n_an_bn^{a'}n^{b'}n^{c'} \,, \label{S4}
\eeqa
and
\beqa
    {{S^{(\theta)}}_{ab}}^{c'a'b'}
    &=& \frac{22\sqrt{z(1-z)}}{9z^4}\left\{ -2\left(1+\frac{4z}{3}\right) g^{c'a'}{g_{(a}}^{b'}n_{b)}
    - 2{g_{(a}}^{c'}{g_{b)}}^{a'}n^{b'}
    - {g_{(a}}^{a'}{g_{b)}}^{b'}n^{c'} \right. \nol \\
    && + \left(\frac{4z^2}{3}-1\right)g_{ab}g^{c'a'}n^{b'}
    + \left(\frac{8z}{3}-1\right)g^{a'b'}{g_{(a}}^{c'}n_{b)}
    - \left(\frac{4z^2}{3}-2z+\frac{1}{2}\right)g^{a'b'}g_{ab}n^{c'} \nol \\
    && + 2(1-z)\left[ \left(1-\frac{4z}{3}\right)g^{a'b'}n_an_bn^{c'}
    + (2-z) g_{ab}n^{a'}n^{b'}n^{c'} + 2{g_{(a}}^{c'}n_{b)}n^{a'}n^{b'} \right. \nol \\
    && + \left.\left.4{g_{(a}}^{a'}n_{b)}n^{b'}n^{c'}
    + 2\left(1+\frac{2z}{3}\right) g^{c'a'}n_an_bn^{b'} \right]
    - 4(1-z)^2n_an_bn^{a'}n^{b'}n^{c'} \right\} \,. \label{S5}
\eeqa
These formulas have been simplified by using
the identities
\beqa
    \left[f(z)-f(1)\right]\delta^{(n)}(1-z) & = &
     \sum^n_{k=1}\left[\frac{(-1)^{k+1}n!}{k!(n-k)!}f^k(z)\delta^{(n-k)}(1-z)\right] \,, \label{Delta_Reduction}\\
    f(z)\delta^{(n)}(1-z) & = & \sum_{k=0}^{n}\left[ \frac{n!}{k!(n-k)!}f^{(k)}(1)\delta^{(n-k)}(1-z) \right] \,. \label{Delta_simplification}
\eeqa

To evaluate the initial field we substitute in Eq.~(\ref{key formula:initial}) the initial data $h_{ab}$ and $\pi^{\lambda'ab}$ given by Eqs.~(\ref{h00})-(\ref{h33}) (with hatted variables) and Eqs.~(\ref{AHpi1})-(\ref{AHpi3}), respectively, and the Green's function $G_{aba'b'}$ given by Eqs.~(\ref{AHGdelta}) and the function
${\left(L_{\pi}G\right)_{ab}}^{\lambda'a'b'}$ given by Eq.~(\ref{AHG}).  In performing the integral over the initial hypersurface it is easier to integrate over the angle $\hat\theta'$ first
and then over $\hat r'$.
For this purpose we re-express the delta function as
\beq
    \delta(1-z) = \frac{2\hat\lambda\hat\lambda'}{\hat r\hat r'} \,
    \delta \left( \cos\hat\theta' - \frac{\hat r^2+\hat r'^2-(\hat\lambda-\hat\lambda')^2}{2\hat r\hat r'} \right) \,.
    \label{AHdelta1-z}
\eeq
This can be used to perform the $\hat\theta'$-integral of the form
$$
\int_0^\infty d\hat r'\int_{-1}^1 d\cos\hat\theta'\,f(\hat r',\cos\hat\theta')\delta^{(n)}(1-z)
$$
for $n=0,1,2,3$, where $f(\hat r',\cos\hat\theta')$ is any function of $\hat r'$ and $\cos\hat\theta'$.  The outline of the lengthy calculation together with some intermediate formulas used are given in Appendix~\ref{AppendixB}.  The resulting initial field is given by
\beqa
    h^{(I)}_{\hat\lambda\hat\lambda}
        &=& \frac{4M}{3H}\left[ \frac{2}{\hat r(\hat r+\hat\lambda)} - \frac{1}{(\hat r+\hat\lambda)^2}
        - \frac{1}{2\hat r\hat\lambda} \right]\theta(\hat r-\hat\lambda+\hat\lambda') \nol \\
        && + \frac{4M}{3H} \left[ \frac{2}{\hat r(\hat r+\hat\lambda)}
        - \frac{1}{(\hat r+\hat\lambda)^2} + \frac{2}{\hat r(\hat r-\hat\lambda)}
        - \frac{1}{(\hat r-\hat\lambda)^2} + \frac{3}{\hat\lambda^2} \right]\theta(\hat\lambda-\hat\lambda'-\hat r) \,, \\
    h^{(I)}_{\hat r\hat r} &=& \frac{4M}{3H} \left[ \frac{2}{\hat r(\hat r+\hat\lambda)} - \frac{1}{(\hat r+\hat\lambda)^2}
        + \frac{2}{\hat\lambda^2} - \frac{1}{2\hat r\hat\lambda} \right]\theta(\hat r-\hat\lambda+\hat\lambda') \nol \\
        && +\frac{4M}{3H}\left[ \frac{2}{\hat r(\hat r+\hat\lambda)} - \frac{1}{(\hat r+\hat\lambda)^2}
        + \frac{2}{\hat r(\hat r-\hat\lambda)}
        - \frac{1}{(\hat r-\hat\lambda)^2} + \frac{5}{\hat\lambda^2} \right]\theta(\hat\lambda-\hat\lambda'-\hat r) \,, \\
    h^{(I)}_{\hat\lambda \hat r} 
    &=& \frac{4M}{3H} \left[ \frac{2}{\hat r(\hat r+\hat\lambda)}
        - \frac{1}{(\hat r+\hat\lambda)^2} - \frac{2}{\hat r\hat\lambda} \right]\theta(\hat r-\hat\lambda+\hat\lambda') \nol \\
        && + \frac{4M}{3H}\left[ \frac{2}{\hat r(\hat r+\hat\lambda)} - \frac{1}{(\hat r+\hat\lambda)^2}
        - \frac{2}{\hat r(\hat r-\hat\lambda)}
        + \frac{1}{(\hat r-\hat\lambda)^2}
        - \frac{4}{\hat r\hat\lambda} - \frac{3\hat r}{2\hat\lambda^3} \right] \nonumber \\
        && \times\theta(\hat\lambda-\hat\lambda'-\hat r) \,, \\
    h^{(I)}_{ij}
        &=& -\frac{4M}{3H} \left[ \frac{\hat r}{2\hat\lambda} + \frac{\hat r^2}{\hat\lambda^2} \right]
        \theta(\hat r-\hat\lambda+\hat\lambda')
        -\frac{4M}{3H}\frac{\hat r^2}{\hat\lambda^2}\theta(\hat\lambda-\hat\lambda'-\hat r)\eta_{ij} \,.
\eeqa
We see that the input field on the initial hypersurface with the time variable $\hat\lambda'$ is reproduced for $\hat{r}>\hat\lambda-\hat\lambda'$, i.e.\ in the region which is not in the causal future of either of the two mass points.

As we observed before, the initial Cauchy surface coincides with past infinity in the limit $\hat\lambda'\to 0$.  In this limit the terms proportional to $\theta(\hat\lambda-\hat\lambda'-\hat r) = \theta(\hat\lambda-\hat r)$ give the initial field in the overlap region covered by both the upper- and lower-half conformally-flat coordinate systems.  By transforming the initial field in this region from the lower-half conformally-flat coordinates system to the upper-half one by Eqs.~(\ref{AHtrans1}) and (\ref{AHtrans2}), we find
\beqa
    \left. h^{(I)}_{\lambda\lambda}\right|_{\rm overlap}
        &=& -\frac{4M}{3H}\,\frac{3}{\lambda^2} \,, \\
    \left. h^{(I)}_{rr}\right|_{\rm overlap} &=& -\frac{4M}{3H}\,\frac{1}{\lambda^2} \,, \\
    \left. h^{(I)}_{\lambda r}\right|_{\rm overlap}  &=& \frac{4M}{3H}\,\frac{3r}{2\lambda^3}\,, \\
    \left. h^{(I)}_{ij}\right|_{\rm overlap}
        &=& -\frac{4M}{3H}\frac{r^2}{\lambda^2}\eta_{ij} \,,
\eeqa
where $i,j = \theta,\phi$.  By comparing these formulas with the source field given by Eq.~(\ref{h00})-(\ref{h33}) and recalling that the initial field is computed with $\alpha=\frac{5}{3}$, we find that the extra contribution in the source field $h_{ab}^{(S)}$ is canceled exactly by the initial field $h^{(I)}_{ab}$.  Thus, the field $h_{ab} = h_{ab}^{(S)}+h_{ab}^{(I)}$ agrees with the input field for $\alpha=\frac{5}{3}$ in the causal past of the North Pole.  The calculation of the field in the causal past of the South Pole is exactly the same.  Hence, by the same argument as in the electromagnetic case, we see that the field $h_{ab}$ is correctly reproduced by the retarded Green's function through the formula (\ref{key formula:gravity}).

\section{Summary}

In this paper we showed in examples that the retarded Green's function does reproduce the electromagnetic and gravitational fields in de~Sitter spacetime through the formula (\ref{key formula}).  This verification is significant because it illustrates the fact that the de~Sitter invariant construction of the retarded Green's function is applicable to generate the fields which satisfy the equations of motion.
Thus, the retarded Green's function should not be abandoned contrary to recent claims.  Our calculations also serve as a consistency check for the propagators for electromagnetism~\cite{AllenJacobson} and linearized gravity~\cite{Higuchi2} in de~Sitter spacetime.

It is interesting that the time component of the electromagnetic field coming from the charge at the North Pole in de~Sitter spacetime possesses an extra term $1/3\lambda$ which does not satisfy the Lorenz gauge condition.
This term is canceled out by the initial-data contribution in the overlapping region of the spacetime (and thus the Lorenz gauge condition is satisfied).
This result illustrates the need for including the initial data on past infinity satisfying the Gauss constraint equation in a spacetime with spacelike past infinity. (The field at a point which is not in the causal future of either charge is reproduced exactly by the field from the initial data alone.)

The way the linearized gravitational field is reproduced by Eq.~(\ref{key formula}) for two mass points is quite similar though the calculation is much more complicated.  Again, the contribution from the source does not quite reproduce the field satisfying the appropriate gauge condition except in the Landau-like gauge ($\alpha=0$) though it does satisfy the linearized Einstein equations for all $\alpha$. We verified that the contributions from the source and initial data together give the correct field in the causal future of the mass points and that the field at a point not in the causal future of the mass points is reproduced by the field from the initial data alone.

\acknowledgments

We are indebted to Demian Cho, who inspired us to carry out this work and contributed to initial discussions.  We are also grateful to Emil Mottola for useful correspondence.  We used Maple 7 in some of our calculations.

\appendix

\section{A proof of Eq.~(\ref{AHGreen})} \label{AppendixA}

The definition of the delta function (\ref{AHdefdelta}) is equivalent to the property
\beq
\int d^4x\sqrt{-g}\,\int d^4 x'\sqrt{-g'}
B_I(x){\delta^I}_{I'}\delta^4(x,x')A^{I'}(x') = \int d^4 x\sqrt{-g}\,B_I(x)A^I(x)
\eeq
for any smooth compactly-supported functions $B^I(x)$ and $A^I(x)$. This implies that the delta function satisfies
\beq
{\delta_{I'}}^I\delta^4(x',x) = {\delta^I}_{I'}\delta^4(x,x')\,.
\eeq
Now, for any smooth compactly-supported function $A_I(x)$ we find by Eq.~(\ref{AHGreen})
\beq
L_x^{IJ}\int d^4x'\sqrt{-g'}\,G^R_{JI'}(x,x')L_{x'}^{I'J'}A_{J'}(x') = L_x^{IJ}A_J(x)\,. \label{AHalmost}
\eeq
Let $B^I(x)$ be an arbitrary smooth compactly-supported function and let $f_I(x)$ be the {\em advanced} solution to the equation $L^{IJ}f_J = B^I$.  Multiplying Eq.~(\ref{AHalmost}) by $f_I(x)$ and integrating over $x$ we have
\beq
\int d^4x\sqrt{-g}\,\int d^4 x'\sqrt{-g'}\,f_I(x)L_x^{IJ}G^R_{JI'}(x,x')L_{x'}^{I'J'}A_{J'}(x') = \int d^4x\sqrt{-g}\,f_I(x)L_x^{IJ}A_J(x)\,.
\eeq
We can integrate by parts in $x$ and $x'$ so that $L_x^{IJ}$ and $L_{x'}^{I'J'}$ act on $f_J(x)$ and $G^R_{JJ'}(x,x')$, respectively, with the following result:
\beq
\int d^4x \sqrt{-g}\,B^I(x)\left[L_{x'}^{I'J'}G^R_{JJ'}(x,x')\right]A_{I'}(x') = \int d^4 x\sqrt{-g}B^I(x)A_I(x)\,.
\eeq
This equation is equivalent to Eq.~(\ref{AHGreen}).

\section{Some formulas used in Sec.~\ref{case3}} \label{AppendixB}

{}From Eq.~(\ref{n}) we find, using  Eqs.~(\ref{AHAJ1})-(\ref{AHAJ4}),
\beqa
    \nabla_a\left(\partial_b\cos H\mu\right) &=& H^2(1-2z)g_{ab} \,, \\
    \nabla_a\left(\partial_{b'}\cos H\mu\right) &=& H^2[g_{ab'}+2(1-z)n_an_{b'}] \,,
\eeqa
which enable us to calculate the derivatives of $C^{(i)}_{aba'b'}$ and $C^{(i)}_{aa'bb'}$, $i=1,2,3$, defined by Eqs.~(\ref{AHC1})-(\ref{AHC3}) as
\beqa
    \nabla^{c'}{{C^{(1)}}_{ab}}^{a'b'} &=& 0 \,, \\
    \nabla^{c'}\({C^{(1)}}_{a\,\,\,\,b}^{\,\,\,a'\,\,\,b'} + {C^{(1)}}^{a'\,\,\,b'}_{\,\,\,\,b\,\,\,\,a}\)
        &=& 4H\frac{\sqrt{z(1-z)}}{z}\[g^{c'(a'}{g_{(a}}^{b')}n_{b)} + {g_{(a}}^{c'}{g_{b)}}^{(a'}n^{b')}\] \,, \\
    \nabla^{c'}{{C^{(2)}}_{ab}}^{a'b'} &=& 4H\sqrt{z(1-z)} \left\{ 2(z-1)g^{a'b'}n_{a}n_{b}n^{c'} \right. \nol \\
        && \qquad + \left. (2z-1)g_{ab}g^{c'(a'}n^{b')} - g^{a'b'}{g_{(a}}^{c'}n_{b)} \right\} \,, \\
    \nabla^{c'}\({C^{(2)}}_{a\,\,\,\,b}^{\,\,\,b'\,\,\,a'} + {C^{(2)}}_{a\,\,\,\,b}^{\,\,\,a'\,\,\,b'}\)
        &=& 8H\sqrt{z(1-z)} \left\{ 2(1-z)\[g^{c'(a'}n_{a}n_{b}n^{b')} + {g_{(a}}^{c'}n^{(a'}n_{b)}n^{b')} \right.\right. \nol \\
            && \qquad - \left.\left.{g_{(a}}^{(a'}n_{b)}n^{b')}n^{c'}\]
            \right.\nonumber \\
            && \qquad \left. - {g_{(a}}^{c'}{g_{b)}}^{(a'}n^{b')}+ (2z-1){g_{(a}}^{(a'}n_{b)}g^{b')c'} \right\} \,, \qquad \\
    \nabla^{c'}{{C^{(3)}}_{ab}}^{a'b'} &=& 16Hz(1-z)\sqrt{z(1-z)}\nonumber \\
    && \times \left\{  (2z-1)g^{c'(a'}n_an_bn^{b')}
         \right. \nol \\
        &&  \qquad \left. - {g_{(a}}^{c'}n_{b)}n^{a'}n^{b'}-2(1-z)n_an_bn^{a'}n^{b'}n^{c'} \right\} \,.
\eeqa
These are used to find ${S^{(i)}_{ab}}^{c'a'b'}$ given by Eqs.~(\ref{S1})-(\ref{S5}).

For any integrand $f(\hat r',\cos\hat\theta')$ , Eqs.~(\ref{Delta_Reduction})-(\ref{AHdelta1-z}) can be used to carry out the $\hat\theta'$-integral.  We define
\beq
    \xi \equiv \frac{\hat r^2+\hat r'^2-(\hat\lambda-\hat\lambda')^2}{2\hat r\hat r'} \,.
\eeq
Then,
\beqa
    && \int_0^{\infty}d\hat r'\int_{-1}^1 d\cos\hat\theta' \, f(\hat r',\cos\hat\theta')\delta(1-z)\nol\\
        &&= \int_{\hat r'=\epsilon[\hat r-(\hat\lambda-\hat\lambda')]}^{\,\hat r+(\hat\lambda-\hat\lambda')}
        d\hat r'\,\frac{2\hat\lambda\hat\lambda'}{\hat r\hat r'}\, f(\hat r',\xi) \,,
\eeqa
\beqa
    && \int_0^{\infty}d\hat r'\int_{-1}^1 d\cos\hat\theta' \, f(\hat r',\cos\hat\theta')\delta'(1-z)\nol\\
        &&= \left. \int_{\hat r'=\epsilon[\hat r-(\hat\lambda-\hat\lambda')]}^{\,\hat r+(\hat\lambda-\hat\lambda')}
        d\hat r'\left(\frac{2\hat\lambda\hat\lambda'}{\hat r\hat r'}\right)^2 \frac{\partial
        f(\hat r',\cos\hat\theta')}{\partial\cos\hat\theta'}
        \right|_{\cos\hat\theta'=\xi} \nol \\
        && \quad - \int_{0}^{\infty} d\hat r'\,\frac{2\hat\lambda\hat\lambda'}{\hat r\hat r'}\,
        \left[ f(\hat r',\cos\hat \theta')\delta(1-z) \right]^1_{\cos\hat \theta'=-1} \,, \qquad
\eeqa
\beqa
&&    \int_0^{\infty}d\hat r'\int_{-1}^1 d\cos\hat\theta' \, f(\hat r',\cos\hat\theta')\delta''(1-z)\nol\\
        &&= \int^{\hat r+(\hat\lambda-\hat\lambda')}_{\epsilon[
        \hat r-(\hat\lambda-\hat\lambda')]}d\hat r'\left(\frac{2\hat\lambda\hat\lambda'}{\hat r\hat r'}\right)^3
        \left.\frac{\partial^2f(\hat r',\cos\hat\theta')}{\partial\cos\hat\theta'^2}\right|_{\cos\hat\theta'=\xi} \nol \\
        && \quad - \int_0^{\infty}d\hat r' \left\{ \,\frac{2\hat\lambda\hat\lambda'}{\hat r\hat r'}\,
        \left[f(\hat r',\cos\hat\theta')\delta'(1-z)\right]^1_{\cos\hat\theta'=-1} \right. \nol \\
        && \quad + \left. \left(\frac{2\hat\lambda\hat\lambda'}{\hat r\hat r'}\right)^2
        \left[ \frac{
        \partial f(\hat r',\cos\hat\theta')}{\partial\cos\hat\theta'} \delta(1-z)\right]^1_{\cos\hat\theta'=-1}\right\} \,,
\eeqa
\beqa
    && \int_0^{\infty}d\hat r'\int_{-1}^1 d\cos\hat\theta' \, f(\hat r',\cos\hat\theta')\delta'''(1-z)\nol\\
        && = \int^{\hat r+(\hat\lambda-\hat\lambda')}_{\epsilon[
        \hat r-(\hat\lambda-\hat\lambda')]}d\hat r'\left(\frac{2\hat\lambda\hat\lambda'}{\hat r\hat r'}\right)^4
        \left.\frac{\partial^3f(\hat r',\cos\hat\theta')}{\partial\cos\hat\theta'^3}\right|_{\cos\hat\theta'=\xi} \nol \\
        && \quad - \int_0^{\infty}d\hat r'\left\{\frac{2\hat\lambda\hat\lambda'}{\hat r\hat r'}\,
        \left[f(\hat r',\cos\hat\theta')\delta''(1-z)\right]^1_{\cos\hat\theta'=-1} \right. \nol \\
        && \quad + \left(\frac{2\hat\lambda\hat\lambda'}{\hat r\hat r'}\right)^2
        \left[\frac{\partial f(\hat r',\cos\hat\theta')}{\partial\cos\hat\theta'}
        \delta'(1-z)\right]^1_{\cos\hat\theta'=-1} \nol \\
        && \quad + \left. \left(\frac{2\hat\lambda\hat\lambda'}{\hat r\hat r'}\right)^3
        \left[\frac{\partial^2f(\hat r',\cos\hat\theta')}
        {\partial\cos\hat\theta'^2} \delta(1-z)\right]^1_{\cos\hat\theta'=-1} \right\} \,,
\eeqa
where we use $\epsilon=-1$ ($\epsilon=1$) to calculate the field in the region which is (not) in the causal future of the mass point,
and
\beqa
    && \int_0^{\infty}d\hat r'\int_{-1}^1 d\cos\hat\theta' \, f(\hat r',\cos\hat\theta')\theta(z-1) \nol \\
        &&=
        \begin{cases}
        \displaystyle \int_{\hat r-(\hat\lambda-\hat\lambda')}^{\hat r+(\hat\lambda-\hat\lambda')}
        d\hat r'\int_{\xi}^1 d\cos\hat\theta' \, f(\hat r',\cos\hat\theta')\,, &
        \mbox{if $\hat r>\hat\lambda-\hat\lambda'$} \,; \nol \\
        \displaystyle \left( \int_{-\hat r+(\hat\lambda-\hat\lambda')}^{\hat r+(\hat\lambda-\hat\lambda')}
        d\hat r'\int_{\xi}^1 d\cos\hat\theta'
        + \int_0^{-\hat r+(\hat\lambda-\hat\lambda')}d\hat r'\int_{-1}^1 d\cos\hat\theta' \right)
         \, f(\hat r',\cos\hat\theta')\,, &
        \mbox{if $\hat r<\hat\lambda-\hat\lambda'$} \,.
        \end{cases}\nonumber \\
\eeqa
The expression of the form $\left[F(\hat r',\cos\hat\theta')\delta^{(n)}(1-z)\right]_{\cos\hat\theta'=-1}^1$ appearing in these formulas are evaluated as follows:
\beqa
    && \left[F(\hat r',\cos\hat\theta')\delta(1-z)\right]^1_{\cos\hat\theta'=-1} \nol \\
    && = F(\hat r',\xi)\frac{2\hat\lambda\hat\lambda'}{\hat\lambda-\hat\lambda'}\nol\\
    && \quad \times
    \left[ \delta(\hat r'-\hat r-(\hat\lambda-\hat\lambda')) + \delta(\hat r'-\hat r+(\hat\lambda-\hat\lambda'))
    - \delta(\hat r'+\hat r-(\hat\lambda-\hat\lambda')) \right] \,,
\eeqa
\beqa
    && \left[F(\hat r',\cos\hat\theta')\delta'(1-z)\right]^1_{\cos\hat\theta'=-1} \nol \\
    && \qquad = -F(\hat r',1)\frac{2\hat\lambda\hat\lambda'}{\hat r-\hat r'}\frac{\partial}{\partial \hat r'}
    \left\{ \frac{2\hat\lambda\hat\lambda'}
    {\hat\lambda-\hat\lambda'}[\delta(\hat r'-\hat r-(\hat\lambda-\hat\lambda'))+
    \delta(\hat r'-\hat r+(\hat\lambda-\hat\lambda'))] \right\} \nol \\
    && \qquad \quad -F(\hat r',-1)\frac{2\hat\lambda\hat\lambda'}{\hat r+\hat r'}\frac{\partial}{\partial \hat r'}
    \left\{ \frac{2\hat\lambda\hat\lambda'}{\hat\lambda-\hat\lambda'}
    \delta(\hat r'+\hat r-(\hat\lambda-\hat\lambda'))\right\} \,,
\eeqa
and
\beqa
    && \left[F(\hat r',\cos\hat\theta')\delta''(1-z)\right]^1_{\cos\hat\theta'=-1} \nol \\
    && = F(\hat r',1)\left[ \frac{(2\hat\lambda\hat\lambda')^2}{(\hat r-\hat r')^3}
    \frac{\partial}{\partial \hat r'}
    + \left(\frac{2\hat\lambda\hat\lambda'}{\hat r-\hat r'}\right)^2\frac{\partial^2}{\partial \hat r'^2} \right]
    \frac{2\hat\lambda\hat\lambda'}{\hat\lambda-\hat\lambda'}\nol\\
    && \times
    \[\delta(\hat r'-\hat r-(\hat\lambda-\hat\lambda'))+\delta(\hat r'-\hat r+(\hat\lambda-\hat\lambda'))\] \nol \\
    && \quad + F(\hat r',-1)\left[ \frac{(2\hat\lambda\hat\lambda')^2}{(\hat r+\hat r')^3}
    \frac{\partial}{\partial \hat r'}
    - \left(\frac{2\hat\lambda\hat\lambda'}{\hat r+\hat r'}\right)^2\frac{\partial^2}{\partial \hat r'^2} \right]
    \frac{2\hat\lambda\hat\lambda'}{\hat\lambda-\hat\lambda'}\delta(\hat r'+\hat r-(\hat\lambda-\hat\lambda')) \,.
\eeqa

\end{document}